\documentclass[%
 aps,
 pre,
 amsmath,amssymb,
reprint,%
longbibliography]{revtex4-2}
\usepackage{stackengine}

\usepackage[export]{adjustbox}
\usepackage{subcaption}
\usepackage{mwe}
\usepackage{ragged2e}

\usepackage{amsmath}
\usepackage{courier}
\usepackage{amssymb}
\usepackage{amsthm}
\usepackage[T1]{fontenc}
\usepackage[latin9]{inputenc}
\usepackage{graphicx}        
\usepackage[bottom]{footmisc}
\usepackage{mathtools}        
\usepackage{comment}

\usepackage{hyperref}
\hypersetup{colorlinks=true,
  linkcolor=blue,
  anchorcolor=blue,
  citecolor=red,
  urlcolor=blue,
  pdfauthor={Antonio Moro}
}

\providecommand{\be}{\begin{equation}}
\providecommand{\ee}{\end{equation}}
\providecommand{\bea}{\begin{eqnarray}}
\providecommand{\eea}{\end{eqnarray}}
\providecommand{\beas}{\begin{eqnarray*}}
\providecommand{\eeas}{\end{eqnarray*}}

\providecommand{\beni}{\begin{equation*}}
\providecommand{\eeni}{\end{equation*}}

\providecommand{\bw}{\begin{widetext}}
\providecommand{\ew}{\end{widetext}}

\def\be{\begin{equation}}
\def\ee{\end{equation}}
\def\bfig{\begin{figure}[htb]}
\def\efig{\end{figure}}

\newcommand{\benumerate}{\begin{enumerate}}
\newcommand{\eenumerate}{\end{enumerate}}

\newcommand{\der}[2]{\frac{\partial #1}{\partial #2}}

\newcommand{\dersec}[2]{\frac{\partial^{2} #1}{\partial #2^{2}}}

\newcommand{\derk}[2]{\frac{\partial^{k} #1}{\partial #2^{k}}}

\newcommand{\av}[1]{\langle #1 \rangle}

\begin{document}




\title{$p$-star models, mean field random networks and the heat hierarchy}


\author{Gino Biondini$^{a)}$, Antonio Moro$^{b)}$, Barbara Prinari$^{a)}$, Oleg Senkevich$^{b)}$}

\address{$^{a)}$ Department of Mathematics University at Buffalo, State University of New York, USA \\
$^{b)}$Department of Mathematics, Physics and Electrical Engineering, Northumbria University Newcastle, United Kingdom }










\begin{abstract}
We consider the mean field analog of the $p$-star model for homogeneous random networks, and compare its behaviour with that of the $p$-star model and its classical mean field approximation in the thermodynamic regime. We show that the partition function of the mean field model satisfies a sequence of partial differential equations known as the heat hierarchy, and the models connectance is obtained as a solution of a hierarchy of nonlinear viscous PDEs. In the thermodynamic limit, the leading order solution develops singularities in the space of parameters that evolve as classical shocks regularised by a viscous term. Shocks are associated with phase transitions and stable states are automatically selected consistently with the Maxwell construction. The case $p=3$ is studied in detail. Monte Carlo simulations show an excellent agreement between the $p$-star model and its mean field analog at the macroscopic level, although significant discrepancies arise when local features are compared. \\

\noindent Keywords: Exponential Random Networks $|$ Mean Field Network Models  $|$ Heat Hierarchy $|$ Integrability

\end{abstract}

\maketitle

\section{Introduction} Networks provide an effective conceptual framework to model complex systems where fundamental constituents and interactions can be represented respectively by nodes and links \cite{Newman_book, Barabasi_book}. Following a standard mathematical terminology, a network is a graph where vertices correspond to nodes and links to edges. Graph theory, introduced by Leonard Euler in 1736 and inspired by the celebrated problem of the seven bridges of K\"onisberg, has further developed into an established field of mathematics with numerous applications in a variety of disciplines, such as physics, technology and information sciences,  biology, sociology, epidemiology \cite{Newman_book,Barabasi_Network_Biology, Stein_Marks_Sander, Bassett2017,ROBINS2007173, ROBINS2007192, ANDERSON199937, WassPatt1996,Pastor2015}. However, understanding real-world networks is particularly challenging, since they exhibit specific features of complex systems, like for instance absence of equilibrium, complex intrinsic topology, geometry and dynamics, which make detailed analysis and prediction of their behaviour a task currently out of reach. 

Random graph models are often introduced with the aim to capture and provide a qualitative description of macroscopic features of complex networks which may arise independently of the specific microscopic detail of their realisation.
An important class of such models are the Exponential Random Graphs Models (ERGMs). ERGMs are specified by a probability distribution that maximizes the Gibbs entropy subject to constraints on expectation values of observables \cite{Newman_book, ParkNewman2014}. A special family of  ERGMs is represented by networks where only links with a common node interact. These models are referred to as $p$-star models. The Gibbs-Boltzmann (GB) probability distribution of $p$-star models is  of the form~\cite{Newman_book}
\begin{equation} \label{GB_distribution}
P[\mathbf A] = \frac{\mathrm{e}^{-H_p[\mathbf A]}}{Z_n}
\end{equation}
where
\begin{equation}
\label{Hstar}
H_p[\mathbf A] = -\sum_{k=1}^{p} \frac{\tau_{k}}{n^{k-1}}S_k(\mathbf A)
\end{equation}
is the graph Hamiltonian,  $p\in \mathbb N$, $\mathbf A = (A_{ij})_{i,j=1,\dots,n}$ is the $n \times n$ adjacency matrix of a simple undirected graph, i.e. $A_{ij} \in \{0,1\}$ if $i\neq j$ and $A_{ii} = 0$,  $\tau_{k}$ are the coupling constants and
\begin{equation}\label{stars_def}
  S_k(\mathbf A) = \frac{1}{k!}\sum_{i, j_{l} \neq j_{m}} A_{ij_{1}} A_{ij_{2}} \dots A_{ij_{k}}
\end{equation}
is the number of $k$-stars (sets of $k$ links attached to the same node). The partition function is defined in a standard way as
\begin{equation}
\label{Zdef}
Z_{n} = \sum_{\{\mathbf A\}} \mathrm{e}^{-H_p[\mathbf A]},
\end{equation}
where the sum is evaluated over the set $\{\mathbf{A} \}$ all possible configurations of $\mathbf A$. 

In this paper we consider a mean field (MF) analog of the $p$-star model \eqref{Hstar} defined by the  Hamiltonian 
\begin{gather}
\label{Hamp}
\begin{aligned}
H_{\mathrm{MF}}[\mathbf A] &= -\sum_{k=1}^{p} \frac{t_{k}}{(2N)^{k-1}} \Big(\sum_{i,j} A_{ij} \Big)^{k},
\end{aligned}
\end{gather}
which we refer to as MF model.
where $N=n(n-1)/2$ is the maximum number of links in a simple undirected graph of $n$ nodes. The MF model~(\ref{Hamp}) is defined for a finite size system and its applicability relies on the assumption that all $k$-tuples of links interact via effective coupling constant $t_{k}$. The Hamiltonian \eqref{Hamp}, unlike \eqref{Hstar},  accounts indeed for contributions from all interactions of $k$-links for $k\leq p$ independently on whether links share a node or not, and hence is linearly extensive for large $N$, i.e. $H_{\mathrm{MF}}[\mathbf{A}] = O(N)$ as $N \to \infty$ . We compare the qualitative features of both models \eqref{Hstar} and \eqref{Hamp} at finite $n$ and in the thermodynamic limit (i.e. $n \to \infty$).  Using the approach introduced in \cite{BarraMoro2014},  we provide a detailed study of the solvability and the integrability of the mean field model \eqref{Hamp} for any degree of interaction as a solution of the heat hierarchy. The properties of the heat hierarchy and its generalisations from the point of view of the theory of integrable perturbations of the type considered in this paper have been extensively studied in~\cite{ALM2015, ALM2015b}.

For the purposes of this work it is crucial to write the Hamiltonian \eqref{Hamp} in terms of the connectance $L = \sum_{ij} A_{ij}/(2 N)$, that is
\begin{equation}
\label{HamL}
H_{\mathrm{MF}}[L] = -\sum_{k=1}^{p} (2N) t_{k} \; L^{k},
\end{equation}
where the notation emphasises the fact that the Hamiltonian only depends on the connectance of the given configuration.
We demonstrate that the partition function associated to \eqref{HamL} satisfies a compatible hierarchy of linear PDEs (the heat hierarchy), of which the heat equation is the first member and in which the coupling constants $t_{k}$ play the role of independent variables. The large $n$ limit is singular and, with a suitable re-scaling, at the leading order, the expectation value of the connectance satisfies a hierarchy of quasilinear PDEs (the Hopf hierarchy). Solutions to the Hopf hierarchy develop singularity for finite values of the independent variables $t_{k}$ and such singularities are associated to critical points and phase transitions in the system. A comparison between the exact solution of the heat hierarchy and Monte Carlo simulations of both $p$-star models and their mean field analogue shows that the analytical solution provides an accurate quantitative description of the system for sufficiently large $n$.

The paper is organised as follows: in section \ref{sec:background} we show that the partition function for the MF model satisfies the heat hierarchy, and the free energy also satisfies a set of nonlinear viscous PDEs known as the Burgers' hierarchy. In section \ref{sec:MFNetwork} we compare MF networks and p-star models in the thermodynamic limit  by minimising the Kullbach-Leibler divergence. In section \ref{sec:pstar} we discuss the properties of the network connectance, show that it satisfies the Hopf hierarchy, whose generic solution develops a gradient catastrophe singularity and explore the critical sector and singularity resolution. In section \ref{sec:MCS} we provide some thermodynamical considerations and discuss Monte Carlo simulations performed to explore the landscape of the free energy and reconstruct the corresponding profile of the connectance. Section \ref{distributions_subsec} is devoted to the comparison of local properties of MF and p-star models. Finally, section \ref{sec:conclusions} offers some concluding remarks.



\section{The heat hierarchy} 
\label{sec:background}
Following the approach outlined in \cite{BarraMoro2014}, we observe that the partition function for the MF model
\begin{equation}
\label{MFpart}
  Z_{n}^{(\mathrm{MF})}(\mathbf t) = \sum_{\{\mathbf A\}} \mathrm{e}^{-H_{\mathrm{MF}}[L]},
\end{equation}
satisfies a set (hierarchy) of linear partial differential equations of the form
\begin{equation}
\label{heathierarchy}
\der{Z_{n}^{(\mathrm{MF})}}{t_{k}} = \nu^{k-1} \derk{Z_{n}^{(\mathrm{MF})}}{t_{1}} \qquad k = 1,2,\dots,p
\end{equation}
where $\nu = (2 N)^{-1}$. This statement can be verified by direct substitution given the specific form of the Hamiltonian \eqref{HamL}. It is straightforward to verify that equations of the hierarchy \eqref{heathierarchy} are compatible, that is 
\begin{equation*}
  \frac{\partial}{\partial t_k}\left(\frac{\partial Z_n^{(\mathrm{MF})}}{\partial t_l}\right) = \frac{\partial}{\partial t_l}\left(\frac{\partial Z_n^{(\mathrm{MF})}}{\partial t_k}\right)\ \ \ \forall l,k\in\{1,\dots,p\}.
\end{equation*}
Equations \eqref{heathierarchy} are referred to as the heat hierarchy, since the first member of the hierarchy is the heat equation 
\[
\der{Z_{n}^{(\mathrm{MF})}}{t_{2}} = \nu \dersec{Z_{n}^{(\mathrm{MF})}}{t_{1}}.
\]
  
The partition function of the MF model is given by the solution of the hierarchy (\ref{heathierarchy}) that matches the initial condition
\begin{equation} \label{init}
  Z_{n,0}^{(\mathrm{MF})}(t_{1}) \equiv Z_{n}^{(\mathrm{MF})}|_{t_{k}=0\ \forall \, k>1} =   \sum_{\{\mathbf A\}} \mathrm{e}^{ t_{1} \sum_{i,j} A_{ij}}.
\end{equation}
We note that Eq. \eqref{init} represents the partition function of the celebrated Erd\"os-R\'enyi (ER) model \cite{ErdosRenyi1959}, with Hamiltonian
\begin{equation}
\label{HER}
H_{ER} =  - t_{1} \sum_{i,j} A_{ij},
\end{equation}
which defines an exponential random graph model for a network of non-interacting links. Hence, the MF model can be viewed as the ``evolution'' in the space of coupling constants of the ER model according to the heat hierarchy.

As the partition function diverges exponentially in the large $n$ limit, in order to study the MF model in the thermodynamic, i.e. in the limit $n \to \infty$, let us introduce the function
\begin{equation}\label{F_def}
{\cal F}_{n} = \nu \log Z_{n}^{(\mathrm{MF})}.
\end{equation}
In a statistical mechanical context one has
\[
{\cal F}_{n} = -\Phi_{n}/(2T),
\]
where $\Phi_{n}$ is the analog of the specific Helmholtz free energy and $T$ is the temperature of the thermodynamic systems. As the temperature plays the role of a scaling factor in the GB distribution, we incorporate it into the coupling constants. Hence, for an effective comparison we can set $T = 1$. With this choice, $\Phi_{n}$ is expressed in terms of the specific internal energy $E_{n}$ and the specific entropy $S_{n}$ as
\[
\Phi_{n} = E_{n} - S_{n}.
\]
For the sake of clarity in the use of terminology, we note that due to the minus sign, the stable equilibrium associated to the minimum of the Helmholtz free energy corresponds to a maximum of the function ${\cal F}_{n}$.

Rewriting the heat hierarchy (\ref{heathierarchy}) in terms of ${\cal F}_{n}$ we obtain the following hierarchy of nonlinear viscous equations, also known as  the Burgers' hierarchy
\begin{equation}
\label{heatbruno}
\der{{\cal F}_{n}}{t_{k}} = \nu^{k} P_{k}\left [ \frac{1}{\nu} \der{{\cal F}_{n}}{t_{1}} \right], \qquad k=1,2,\dots,p
\end{equation}
where $P_{k}[f]$ denotes the Fa\`a di Bruno polynomials which are defined recursively as follows \cite{deBruno}
\[
P_{k+1}[f] := \der{P_{k}[f]}{t_{1}} + f P_{k}[f], \qquad P_{0}[f] :=1.
\]
We observe that, given a solution ${\cal F}_{n}^{(\mathrm{MF})}(\mathbf t)$ of the hierarchy (\ref{heatbruno}),  the derivatives
\begin{equation}
\label{opar}
\der{{\cal F}_{n}}{t_{k}} = \langle L^{k} \rangle_{n}, \ \ \ k = 1,2,\dots,p,
\end{equation}
where
\begin{equation}
\langle L^{k} \rangle_{n} := \frac{\sum_{\{ {\mathbf A} \}} L^{k} \mathrm{e}^{-H}}{Z_{n}^{(\mathrm{MF})}}, 
\end{equation}
provide an effective way of calculating the expectation value of the connectance $ \langle L \rangle_{n}$ and its higher moments  $\langle L^{k} \rangle_{n}$.

Equations for $\langle L \rangle_{n}$ follow from (\ref{heatbruno}) by differentiating both sides w.r.t.  $t_{1}$ and the result leads to the so-called Burgers' hierarchy. For instance, the first two equations of the hierarchy read as
{\small
  \begin{align}  \label{burgersbruno}
    \begin{split}
      &\der{\langle L \rangle_{n}}{t_{2}} = \der{}{t_{1}} \left(\langle L \rangle_{n}^{2} + \nu \der{\langle L \rangle_{n}}{t_{1}} \right); \\
      &\der{\langle L \rangle_{n}}{t_{3}} = \der{}{t_{1}} \left( \langle L \rangle_{n}^{3} + 3 \nu \langle L \rangle_{n} \der{\langle L \rangle_{n}}{t_{1}} + \nu^{2} \dersec{\langle L \rangle_{n}}{t_{1}}  \right).
    \end{split}
  \end{align}
}

To study the system in the thermodynamic limit, i.e. $n \to \infty$ (or equivalently $\nu \to 0 $), we assume that ${\cal F}_{n}$ admits the expansion of the form
\begin{equation}
\label{Fexpansion}
{\cal F}_{n} = F + F_{1} \nu + F_{2} \nu^{2} +  \mathcal O(\nu^{3}).
\end{equation}
A similar expansion holds for the Helmholtz free energy $\Phi_{n}  = \Phi + O(\nu)$.
Substituting the expansion \eqref{Fexpansion} into (\ref{heatbruno}), at the leading order one gets the hierarchy of the Hamilton-Jacobi equations
\begin{equation}
\label{HJeq}
\der{F}{t_{k}} = \left ( \der{F}{t_{1}} \right)^{k} \qquad k=1,2,\dots,p.
\end{equation}


The expansion (\ref{Fexpansion}) implies $\langle L \rangle_{n} = \langle L \rangle + \mathcal O(\nu)$, and in the thermodynamic limit the leading order term $\langle L \rangle$ of the connectance satisfies the following hierarchy of hyperbolic PDEs (the Hopf hierarchy)
\begin{align}
\label{hopf}
&\der{\langle L \rangle}{t_{k}} = \der{\langle L \rangle^{k}}{t_{1}}, \qquad k =1,\dots,p.
\end{align}
We note that the Hopf equation and its hierarchy emerge in connection with a variety of models in statistical thermodynamics, e.g. van der Waals theory \cite{DeNittisMoro2012, MoroAnnals2014} and $p$-spin models~\cite{BarraMoro2014}, and different models are specified by a different initial condition.
The general solution of the hierarchy (\ref{hopf}) is obtained by the method of characteristics and it is implicitly given by the equations
\begin{equation}
\label{hodograph}
\sum_{k=1}^{p} k \langle L \rangle^{k-1} t_{k} = f\left(\langle L \rangle \right),
\end{equation}
where the function $f\left(\langle L \rangle \right)$ is an arbitrary function of its argument specified by the initial condition.

The general solution of (\ref{HJeq}) obtained by integration along characteristics (on which $\langle L\rangle$ is constant) is given by
\begin{equation} \label{F_MF_general}
  F(\mathbf t) = \sum_{k=1}^p \langle L \rangle^kt_k + g(\langle L\rangle),
\end{equation}
where the function $g$ is an arbitrary function of its argument.

In order to specify the arbitrary functions, we need the explicit expression of the initial condition (\ref{init}), that is the partition function of the ER model \eqref{init}. A direct calculation gives
\begin{align*}
  Z_{n,0}^{(ER)}  &= \sum_{\{\mathbf A\}}\mathrm{e}^{2t_1\sum_{i<j}A_{ij}}=\sum_{\{\mathbf A\}}\prod_{i<j}\mathrm{e}^{2t_1A_{ij}}=\\
  &=\prod_{i<j}\sum_{A_{ij}=0}^1\mathrm{e}^{2t_1A_{ij}}=\prod_{i<j}\left(1+\mathrm{e}^{2t_1}\right) = \left( 1+ \mathrm{e}^{2 t_{1}} \right)^N.
\end{align*}
Therefore, the initial condition for ${\cal F}_{n}$ is given by
\begin{equation}\label{initial_FE}
  {\cal F}_{n,0}(t_{1})=\nu\log\left(Z_{n,0}^{(ER)}\right)=\frac{1}{2}\log\left(1+\mathrm{e}^{2t_1}\right)
\end{equation}
from which, using the equation \eqref{opar}, we obtain the initial condition, i.e. the connectance for the non interacting model
\begin{equation} \label{initL}
\langle L \rangle_{n,0} = \der{{\cal F}_{n,0}}{t_{1}} = \frac{\mathrm{e}^{2t_1}}{1+ \mathrm{e}^{2 t_{1}}}.
\end{equation}
Importantly, ${\cal F}_{n,0}$ and $\langle L \rangle_{n,0}$ do not depend on $n$, hence expressions (\ref{initial_FE}) and (\ref{initL}) also give the initial conditions for the hierarchies (\ref{HJeq}) and (\ref{hopf})  respectively. Evaluating  the equations (\ref{hodograph}) and (\ref{F_MF_general}) at $t_{k} = 0$ for $k>1$ we can specify the  functions $f(z)$ and $g(z)$, i.e.
\begin{equation}\label{arbitrary_f}
  f(z) = \frac{1}{2} \log \frac{z}{1-z},
\end{equation}
\begin{equation}\label{arbitrary_g}
  g(z) = - \frac{1}{2}\log\left[z^{z}\left(1-z\right)^{1-z} \right].
\end{equation}
Combining (\ref{hodograph}) and (\ref{arbitrary_f}) we obtain the thermodynamic equation of state for the MF model
\begin{equation}\label{thermodynamic_MF_equation_of_state}
  \langle L\rangle = \frac{1}{2}\left[1+\tanh\left(\sum_{k=1}^pkt_k\langle L\rangle^{k-1}\right)\right].
\end{equation}
Combining (\ref{F_MF_general}) and (\ref{arbitrary_g}) we construct the leading order Helmholtz free energy
\begin{equation}
\label{Hfree}
\Phi = - 2 F
\end{equation}
where $F$  is the solution to the equation (\ref{HJeq}) matching the initial condition (\ref{initial_FE}), i.e.
\begin{equation} \label{F_MF}
    F(\mathbf t) = \sum_{k=1}^p \langle L \rangle^kt_k - \frac{1}{2}\log\left[\langle L\rangle^{\langle L\rangle }\left(1-\langle L\rangle\right)^{1-\langle L\rangle} \right],
\end{equation}
and $\langle L\rangle$ is a function of $\mathbf t$ implicitly defined by (\ref{thermodynamic_MF_equation_of_state}).
The expression \eqref{F_MF} has the natural thermodynamic interpretation where the first term in the r.h.s is proportional to the internal energy $E$, i.e.
\begin{equation}
\label{intenergy}
E = - 2 \sum_{k=1}^p \langle L \rangle^kt_k 
\end{equation}
and the second term is identified with the entropy of the system, that is
\begin{equation} \label{entropy}
  S = -\langle L\rangle\log\langle L\rangle - \left(1-\langle L\rangle\right)\log\left(1-\langle L\rangle\right).
\end{equation}

\section{MF network and $p$-star model}
\label{sec:MFNetwork}
As mentioned above, the MF model~(\ref{Hamp}) is defined for a finite size system and its applicability relies on the assumption that all $k$-tuples of links interact via effective coupling constant $t_{k}$. This is fundamentally different from the $p$-star model \eqref{Hstar} where only $k$-tuples of links sharing one and the same node interact with a coupling constant $\tau_{k}$ and all other $k-$tuples do not interact, i.e. for $k$-tuples that do not share a link  the coupling constant vanishes. We show that the mean field approximation of the $p$-star model, expected to be valid at the equilibrium in the thermodynamic limit, is mapped into the thermodynamic limit of the MF model -  and the solutions to the Hopf hierarchy -  by a rescaling of the coupling constant. 

{\it Mean field approximation.} Let us introduce the probability distribution $P$ associated to the $p$-star model Hamiltonian $H_{p}$ and the probability distribution $Q$ associated to the ER Hamiltonian of the form $H_{\mathrm{ER}} = - h \sum_{i,j} A_{ij}$, i.e.
\begin{equation}
\label{PQdistr}
P = \frac{e^{-H_{p}}}{Z_{p}} \qquad Q = \frac{e^{-H_{ER}}}{Z_{ER}}.
\end{equation}
Gibbs' inequality establishes that \cite{McKay_book}
\begin{equation}
\label{KLdef}
D(h) = \sum_{\{\mathbf{A} \}} Q \log \frac{Q}{P} \geq 0
\end{equation}
where $D(h)$ is referred to as Kullback-Leibler divergence. In particular equality holds only if $P = Q$. As in our case $P\neq Q$, we look for a mean field approximation for the distribution $P$ of the Erd\"os-Reniy form $Q$ as given in \eqref{PQdistr}. $Q$ is chosen  such that $D(h)$ attains its minimum as a function of the variational parameter $h$.
Introducing the free energy in the form
\begin{equation}
\label{KLfree}
\phi(h) = \frac{1}{N} \sum_{\{\mathbf{A} \}} Q \log \frac{Q}{P} - \frac{1}{N} \log Z_{p},
\end{equation}
we observe that, as $Z_{p}$ does not depend on $h$, $D(h)$ and $\phi(h)$ simultaneously attain a minimum as a function of $h$. Substituting the expressions of the probability distributions $P$ and $Q$ defined in \eqref{PQdistr} into the definition of $D(h)$ we have the following identity
\begin{equation}
\label{PQidentity}
\sum_{\{\mathbf{A} \}} Q \log \frac{Q}{P} - \log Z_{p} = \av{H_{p}}_{Q} - \av{H_{ER}}_{Q} - \log Z_{ER}
\end{equation}
where $\av{H_{*}}_{Q}$ is the expectation value of $H_{*}$ w.r.t. the ER distribution $Q$, i.e.
\[
\av{H_{*}}_{Q} = \sum_{\{\mathbf{A} \}} Q H_{*}.
\]
Hence, we can write
\begin{equation}
\label{KLfreeQ}
\phi(h) = \av{H_{p}}_{Q} - \av{H_{ER}}_{Q} - \log Z_{ER}.
\end{equation}
The advantage of the expression \eqref{KLfreeQ} compared with \eqref{KLfree} is that all terms can be evaluated explicitly as the expectation values are calculated w.r.t. the ER probability distribution
\begin{align*}
\av{H_{ER}}_{Q} &= -2 N h \av{L}_{Q}, \\
\av{H_{p}}_{Q} &= -\sum_{k=1}^{p} \frac{\tau_{k}}{n^{k-1}} (n-k) \binom{n}{k}  \av{L}_{Q}^{k},
\end{align*}
where as discussed above for ER model
\begin{align}
\label{ZER_exp}
Z_{ER} &= N \log (1+ e^{2 h}), \\ 
\label{LQ_exp}
\av{L}_{Q} &= \frac{e^{2h}}{1+ e^{2 h}}.
\end{align}
Hence the condition
\[
\der{\phi}{h} = 0
\]
gives
\begin{equation}
\label{hstat}
h = \sum_{k=1}^{p} \frac{k\tau_{k}}{n^{k}} \frac{n-k}{n-1} \binom{n}{k} \left(\frac{e^{2h}}{1+ e^{2 h}} \right)^{k-1}.
\end{equation}
Observing, as it follows from equation~\eqref{LQ_exp}, that
\[
h = \textup{arctanh} (2 \av{L}_{Q}-1),
\]
we obtain the self consistency equation
\begin{equation}
\label{Lvar}
\av{L}_{Q} = \frac{1}{2} \left( 1+ \tanh  \sum_{k=1}^{p} \frac{k\tau_{k}}{n^{k}} \frac{n-k}{n-1} \binom{n}{k} \av{L}_{Q}^{k-1} \right).
\end{equation}
In the thermodynamic regime, direct comparison of the above self-consistency equation \eqref{Lvar} and the equation of state~\eqref{thermodynamic_MF_equation_of_state} lead to the identification
\begin{equation}
\label{ttau}
 t_{k} = \frac{\tau_{k}}{n^{k} } \frac{n-k}{(n-1)}  \binom{n}{k}.
\end{equation}
In particular, we have $t_{1} = \tau_{1}$. Observe that,  for a large network ($n\to \infty$), Eq. \eqref{ttau} implies that $\tau_{k} \simeq k! t_{k}$. Thus, the identification of the $p$-star model Hamiltonian \eqref{Hstar} in the mean field regime with the MF model Hamiltonian \eqref{Hamp}, also implies that $H_{p}[\mathbf{A}] = O(N) = O(n^{2})$ and therefore
\[
S_{k}  = O(n^{k+1}), \qquad n \to \infty.
\]

\section{Critical sector and singularity resolution}
\label{sec:pstar}
{\it Gradient catastrophe.} Based on the relation \eqref{opar}, the expectation value of the connectance $\av{L}$ is calculated as the derivative of the free energy ${\cal F}_{n}$ w.r.t.\ its conjugated variable $t_{1}$, and therefore it is interpreted as an order parameter of the theory. According to the standard statistical mechanical interpretation, singularities of the order parameter are associated to the critical point of  a phase transition in the system.
As demonstrated in the above section \ref{sec:background}, the equation of state \eqref{thermodynamic_MF_equation_of_state}, and similarly \eqref{self_pstar} (and therefore $\av{L}$ as a function of $t_{k}$'s), can be interpreted as a solution of the hierarchy of nonlinear hyperbolic PDEs \eqref{hopf} (the Hopf hierarchy), whose generic solution is known to develop a gradient catastrophe singularity in the $(t_{1},t_{k})$ plane for any given $k>1$ (see e.g. \cite{Whitham_book} and \cite{DeNittisMoro2012}). To obtain the singularity loci in the space of coupling constants $t_{k}$, let us re-write the equation \eqref{hodograph} in the following equivalent form
\begin{equation} \label{t1_thermodynamic_MF_equation_of_state}
\Omega(\av{L}) \equiv t_1 +\sum_{k=2}^pkt_k\langle L\rangle^{k-1} - \frac{1}{2}\log\frac{\langle L\rangle}{1-\langle L\rangle} = 0.
\end{equation}
The conditions that characterise the presence of a cusp singularity are \cite{Whitney1955}
\begin{equation}
\frac{ \partial \Omega}{\partial\langle L\rangle}=0, \qquad \frac{\partial^2 \Omega}{\partial \langle L\rangle^2} = 0,
\end{equation}
which give, respectively,
\begin{align} \label{crit_cond}
  \begin{split}
    &\sum_{k=2}^pk(k-1)t_k\langle L\rangle^{k-1} = \frac{1}{2(1-\langle L\rangle)},\\
    &\sum_{k=3}^pk(k-1)(k-2)t_k\langle L\rangle^{k-1} = \frac{2\langle L\rangle-1}{2(1-\langle L\rangle)^2}.
  \end{split}
\end{align}

\begin{figure}[h]
\begin{center}
\includegraphics[width=8.5cm]{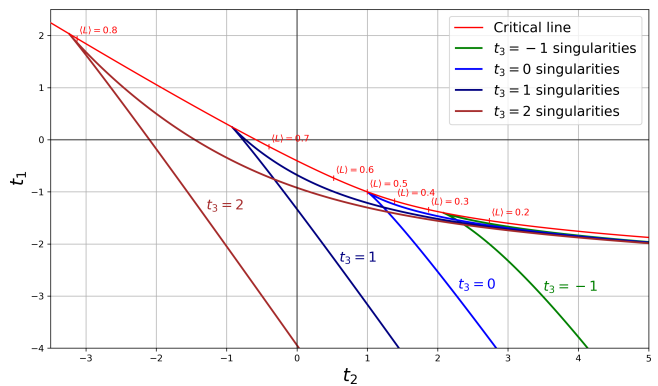} 
\end{center}
\caption{\small {Projection of the critical curve \eqref{crit_line_eqns} on the $(t_{1},t_{2})$ plane. The curve is parametrised by the connectance $\av{L}$ and each point represents the vertex of a sector on the plane where the solution  $\av{L}$  of the corresponding equation of state is multivalued. Examples of sectors are shown for $t_{3} =-1,0,1,2$. }}
\label{fig:critA}
\end{figure}

\begin{figure}[h]
\begin{center}
\includegraphics[width=8.5cm]{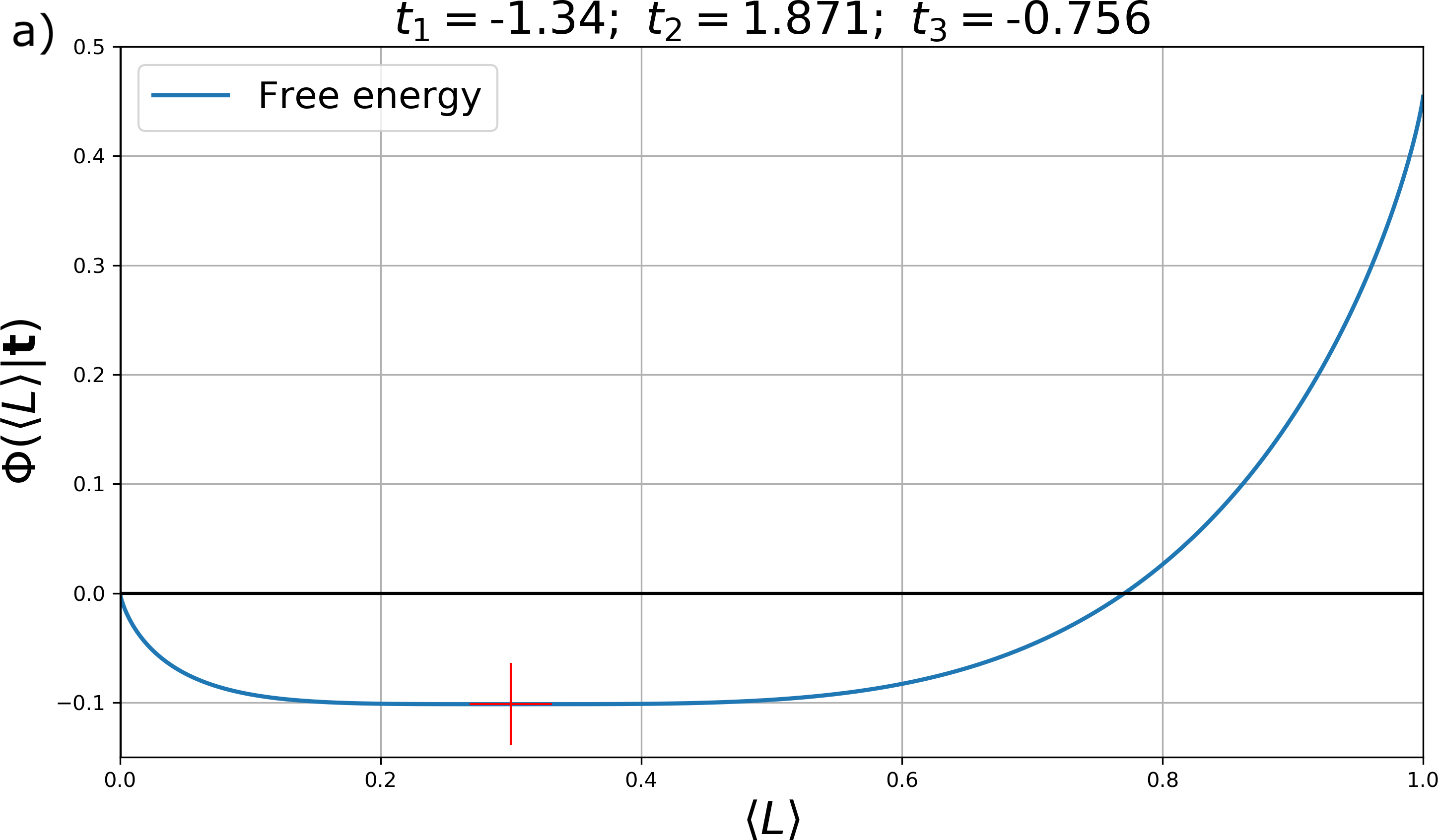} \\
\includegraphics[width=8.5cm]{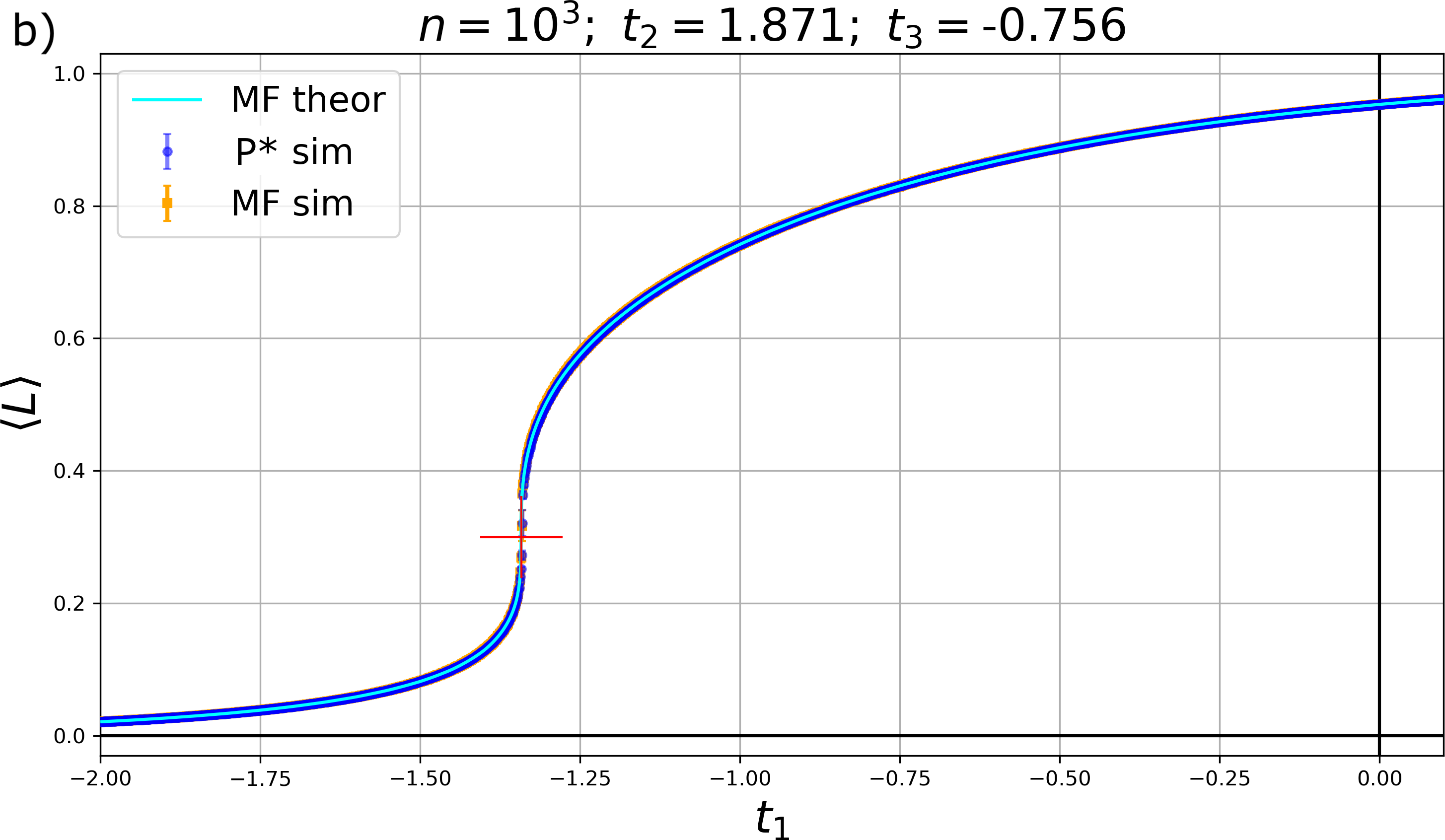} 
\end{center}
\caption{ \small{(a) Profile of the free energy  \eqref{Hfree} in the case $p=3$ for a choice of critical values of coupling constants $t_{1}$, $t_{2}$ and $t_{3}$. The local minimum (highlighted) corresponds to the value of connectance $\av{L}$ at the critical point. (b) Profile of the connectance as a function of $t_{1}$. It shows the occurrence, at the critical point, of a gradient catastrophe singularity. The agreement between the analytical solution of the MF model and MC simulations for both MF and $p$-star (denoted as $p^{*}$ for short) model in the specified regime $n ={10^{3}}$ is compelling.}}
\label{crit_behaviour}
\end{figure}

For example, if $p=2$, Eq.s (\ref{crit_cond}) and (\ref{t1_thermodynamic_MF_equation_of_state}) admits the single point simultaneous solution $(t_{1},t_{2},\av{L}) = (-1,1,1/2)$. For $p>2$ the singular sector is a geometric variety in the space of coupling constants. For instance, in the case $p=3$, the singular sector can be expressed as a curve in the three dimensional space $(t_{1},t_{2},t_{3})$ parametrised by $\av{L}$, that is
\begin{gather}
\label{crit_line_eqns}
\begin{aligned} 
    t_1 &= \frac{1}{2}\log\frac{\langle L\rangle}{1-\langle L\rangle} + \frac{4\langle L\rangle-3}{4(1-\langle L\rangle)^2},\\
    t_2 &= \frac{2-3\langle L\rangle}{4\langle L\rangle(1-\langle L\rangle)^2},\\
    t_3 &= \frac{2\langle L\rangle-1}{12\langle L\rangle^2(1-\langle L\rangle)^2}.
\end{aligned}
\end{gather}
Figure~\ref{fig:critA} shows the projection of the curve \eqref{crit_line_eqns} on  the $(t_1,t_2)$ plane. The curve  separates the region of the plane where the solution $\av{L}$ of the equation of state \eqref{t1_thermodynamic_MF_equation_of_state} is single-valued, i.e. where the free energy admits only one minimum, from the region where the solution $\av{L}$ is multi-valued as a function of coupling constants, i.e. where the free energy admits three critical points. In the latter region the system is bistable, as the analogue of the Helmholtz free energy admits two local minima. Figure \ref{crit_behaviour} shows the profile of the Helmholtz free energy $\Phi = -2  F$, where $F$ is given by the equation \eqref{F_MF}, for a choice of coupling constants $(t_{1},t_{2},t_{3})$ corresponding to the point of gradient catastrophe. The profile of the connectance as a function of $t_{1}$ clearly shows the occurrence of the gradient catastrophe singularity.

{\it Transition formula.} The Hopf hierarchy \eqref{hopf}  provides the leading order asymptotics of the connectance $\av{L}$ as $\nu \to 0$. The gradient catastrophe is associated to the critical point of the phase transition, and multivaluedness captures simultaneous stable and meta-stable states of the system that are associated to multiple local minima of the free energy. In the vicinity of the point of gradient catastrophe, the viscous term represented by the highest derivative in the equation (\ref{heatbruno}) is no longer negligible in the limit $\nu \to 0$. In fact, the viscous contribution, arising from the highest derivatives in \eqref{heatbruno} prevents the gradient catastrophe and the occurrence of multivaluedness \cite{Whitham_book}.

We now construct the solution ${\cal F}_{n}$ to the exact equation (\ref{heatbruno}) and the corresponding connectance $\av{L}= \partial {\cal F}_{n} / \partial t_{1}$ that matches the solution \eqref{thermodynamic_MF_equation_of_state} for $\nu \to 0$ but remains single-valued. For the sake of simplicity we present the explicit calculations for $p=3$.

Let us first observe that the equation (\ref{thermodynamic_MF_equation_of_state}) implies that
\begin{equation}
\label{bcond}
\lim_{t_{1} \to {-\infty}} \langle L \rangle = 0, \qquad \lim_{t_{1} \to {\infty}} \langle L \rangle = 1.
\end{equation}
This is intuitive, as the parameter $t_{1}$ acts as an external field enhancing or preventing the occurrence of links by being largely positive or negative respectively.
We seek a solution that matchs these asymptotic values as the gradient vanishes at $t_{1} \to \pm \infty$.
In particular, we look for the heteroclinic solutions to the system of equations (\ref{burgersbruno}) connecting two equilibrium states at $t_{1} \to \pm \infty$.
Substituting a ``travelling wave'' ansatz of the form
\begin{equation}
\langle L \rangle_{n} = \lambda(\theta) \qquad \theta = t_{1} + c_{2} t_{2} + c_{3} t_{3},
\end{equation}
into \eqref{burgersbruno} and integrating once, one obtains the system of ODEs
\begin{align} \label{lambdaburgers}
  \begin{split}
    &c_{2} \lambda = \lambda^{2} + \nu \lambda' + k_{0} \\
    &c_{3} \lambda = \lambda^{3} + 3 \nu \lambda  \lambda' + \nu^{2} \lambda'' + l_{0},
  \end{split}
\end{align}
where $\lambda' = d \lambda/d\theta$. The boundary conditions~(\ref{bcond}) imply
\[
c_2 = c_3 = 1, \qquad k_0 = l_0 = 0.
\]
The first equation in (\ref{lambdaburgers}) is a separable ordinary partial differential equation which yields
\begin{equation}
  \lambda(\theta) = \frac{1}{2}\left[1+\tanh\left(\frac{1}{2\nu}\left(\theta - c\right)\right)\right],
\end{equation}
and this solution is compatible with the second equation in (\ref{lambdaburgers}). Consistency with the initial condition (\ref{initL}) requires that $\langle L \rangle_n(\mathbf t=\mathbf 0)=1/2$, from which it follows that the integration constant is $c=0$. Finally we obtain the {\it transition formula}
\begin{equation}\label{nushock}
  \langle L\rangle(\mathbf t) = \frac{1}{2}\left[1+\tanh\left(\frac{1}{2\nu}\left(t_1 + t_2 + t_3\right)\right)\right],
\end{equation}
which smoothly connects the asymptotic states specified by the boundary conditions (\ref{bcond}) and describes analytically a shock singularity at $t_{1}+ t_{2} + t_{3} = 0$ in the limit $\nu \to 0$.

{\it $(0,1)$-bistability approximation.} For any choice of coupling constants $t_{k}$ such that the free energy of the $p$-star model admits two sufficiently deep minima at values of $\langle L\rangle$ sufficiently close to $0$ and $1$, as shown in Figure~\ref{FE_standalone} and Figure~\ref{time_avrg_L}, we can make the approximation that the system is found in either of two configurations $A^{C}$, corresponding to the complete graph where all links are active, or $A^{E}$, corresponding to the empty graph where all links are inactive. Under these assumptions we have
\begin{equation} \label{sum_to_one}
  p(\mathbf A^C) + p(\mathbf A^E) \simeq 1,
\end{equation}
where 
\begin{equation} \label{probabilities_C_E}
  p(\mathbf A^C) = \frac{1}{Z}\mathrm{e}^{-H(\mathbf A^C)},\ \  p(\mathbf A^E) = \frac{1}{Z}\mathrm{e}^{-H(\mathbf A^E)}.
\end{equation}
and the partition function is
\begin{equation}
  Z \simeq \mathrm{e}^{-H(\mathbf A^C)} + \mathrm{e}^{-H(\mathbf A^E)}.
\end{equation}
The expectation value of the $k$-star $S_{k}$ is therefore evaluated as
\begin{gather}
\label{Skdeep}
\begin{aligned}
  \langle S_k \rangle \simeq & S_k(\mathbf A^C)\cdot p(\mathbf A^C) + S_k(\mathbf A^E)\cdot p(\mathbf A^E)\\
  \simeq & n\binom{n-1}{k}\cdot \frac{\mathrm{e}^{-H(\mathbf A^C)}}{\mathrm{e}^{-H(\mathbf A^C)} + \mathrm{e}^{-H(\mathbf A^E)}}
\end{aligned}
\end{gather}
where we observed that $S_k(\mathbf A^E) = 0$.
Using the expressions \eqref{probabilities_C_E} and the fact that, given the expression \eqref{Hstar}, $H(\mathbf A^E) = 0$  and
\[
H(\mathbf A^C) = - \sum_{s =1}^{p} \binom{n}{s} \frac{(n-s)}{n^{s-1}}  \tau_s.
\]
The expression~\eqref{Skdeep} gives
{\small
\begin{equation}\label{semiempirical}
  \langle S_k \rangle  \simeq \frac{n}{2}\binom{n-1}{k}\left[1 + \tanh\left(\sum_{s =1}^{p}  \binom{n}{s} \frac{(n-s)}{2 n^{s-1}}  \tau_s\right)\right].
\end{equation}
}
Observing that $\av{L} = \av{S_{1}}/(2N)$, formula \eqref{semiempirical} above gives
\begin{equation}
\label{Lcheck}
\av{L} \simeq \frac{1}{2} \left[1 + \tanh\left(\sum_{k =1}^{p}  \binom{n}{k} \frac{(n-k)}{2 n^{k-1}}  \tau_k\right)\right]
\end{equation}
which, given the identification of the coupling constants \eqref{ttau}, is consistent with the equation \eqref{nushock}. 
We point out that the $(0,1)$-bistability approximation for a $p$-star model with $p>2$ can be achieved only for particular choices of the coupling constants. 

{\it Remark.} It is interesting to compare the mean field approximation of the $p$-star model obtained via the minimisation of the Kullback-Leibler divergence and a common approach (see e.g.\ \cite{Newman_book}) based on the assumption that in each star a link is coupled with the average link $\pi = \av{A_{ij}}$. Therefore, given the $k$th-star interaction term \eqref{stars_def}, namely,
\begin{equation}
\label{stars_def_bis}
S_{k} = \frac{1}{k!} \sum_{i} \sum_{j_{1} } A_{ij_{1}}  \sum_{\substack{j_{2} \neq j_{1} \\ j_{2} \neq i}} A_{ij_{2}} \dots \sum_{\substack{j_{k} \neq j_{l} \\ l = 1,\dots k-1 \\ j_{k} \neq i}} A_{ij_{k}}   
\end{equation}
and replacing $A_{i,j_{s}} \to \av{A_{ij_{s}}}\simeq \pi$ for $s = 2,\dots,k$ for $j_{s} \neq i$
we have
\begin{equation}
\label{star_approx0}
S_{k} \simeq \frac{1}{k!}  (n-2) \dots (n-k) \pi^{k-1} \sum_{i,j}  A_{ij}
\end{equation}
which leads to the Erd\"os-R\'enyi Hamiltonian of the form
\[
H_{p} \simeq H_{ER} = - \bar{t}_{1} \sum_{i,j} A_{i,j},
\] 
with
\[
\bar{t}_{1} = \sum_{k=1}^{p} \frac{\tau_{k}}{n^{k} } \frac{n-k}{(n-1)}  \binom{n}{k} \pi^{k-1}.
\]
Then the solution \eqref{initL} and the hypothesis $\pi = \av{L}$ give the self-consistency equation
\begin{equation}
\label{self_pstar}
\av{L} = \frac{1}{2} \left( 1+ \tanh{\sum_{k=1}^{p} \frac{\tau_{k}}{n^{k} } \frac{n-k}{(n-1)}  \binom{n}{k}  \av{L}^{k-1}} \right),
\end{equation}
leading to the identification
\begin{equation}
\label{ttau2}
 t_{k} = \frac{\tau_{k}}{k n^{k} } \frac{n-k}{(n-1)}  \binom{n}{k}.
\end{equation}
It is immediate to verify that, although the self-consistency equations \eqref{self_pstar} and \eqref{Lvar} are of the same form, Eq. \eqref{Lcheck} is not consistent with \eqref{nushock} under the identification \eqref{ttau2}.
Hence, while the approximation \eqref{star_approx0} predicts the same qualitative behaviours obtained from minimisation of the Kullack-Leibler divergence, the positions of the transition regions do not coincide, as the identification of the parameters \eqref{ttau2} differs by a factor $1/k$ compared with the expression \eqref{ttau}.

\section{Thermodynamic considerations and Monte Carlo simulations} 
\label{sec:MCS}
In this section we use Monte Carlo (MC) simulations to explore the landscape of the free energy \eqref{phiMF}   and reconstruct the corresponding profile of the connectance $\av{L}$. 
\label{sec:montecarlo}
\begin{figure}[t]
\begin{center}
\includegraphics[width=8.5cm]{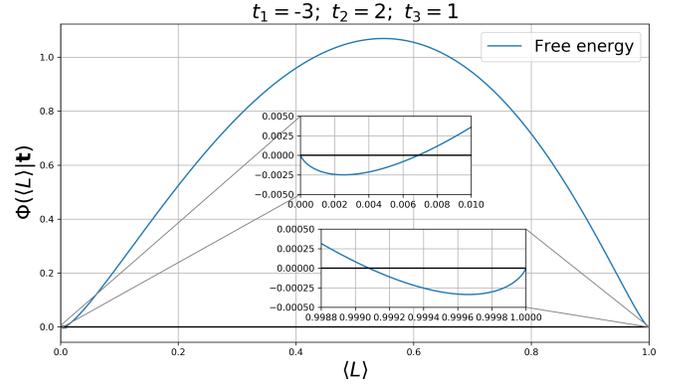}
\end{center}
\caption{\small{Free energy \eqref{phiMF} as a function of connectance for the 3-star model for a choice of parameters such that connectance is a multivalued function of the coupling constants. In this example, the connectance admits two possible values associated to the two minima of almost equal depth located in the vicinity of $\langle L\rangle = 0$ and $\langle L\rangle =1$.}}
\label{FE_standalone}
\end{figure}

\begin{figure}[h]
\begin{center}
\includegraphics[width=8.5cm]{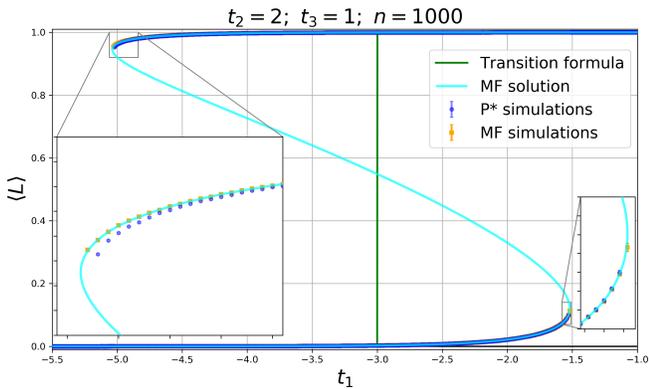}
\end{center}
\caption{\small{Comparison of the solution to the equation of state (\ref{thermodynamic_MF_equation_of_state}) and MC (time averaging) simulations for the corresponding 3-star and MF models with $10^3$ nodes. The theoretical formula  (\ref{thermodynamic_MF_equation_of_state}) is appears accurate for both models apart from small deviations, in the case of $3-$star model, when approaching the multivaluedness region (see magnified in-sets). The branch of the mean field solution on which $\langle L\rangle$ decreases with $t_1$ is unstable due to the negative susceptibility and corresponds to the (local) maximum of the free energy (Figure \ref{FE_standalone}). The green line corresponds to the transition formula (\ref{nushock}).}}
\label{time_avrg_L}
\end{figure}

For the simulations described in this work, we use a Markov chain based method known as Metropolis-Hastings algorithm \cite{metropolis, hastings}. This method is effective in identifying the ground state in the region of parameters where the free energy admits a single minimum. However, performing simulations in the region of coupling constants such that the free energy admits multiple minima requires a careful analysis. Meta-stable states, i.e. local but not absolute minima of the free energy,  behave as attractors, and transitions to the stable state associated to the absolute minimum of the free energy  may require a long iteration time. Indeed, according to the N\'eel relaxation theory~\cite{Fuller1963}, the typical transition time from a  meta-stable to a stable state grows exponentially with the  size of the  energy barrier between the two states, which in turn is proportional to the size $N$ of the system. Therefore, simulations of states fluctuating around a meta-stable state produce a time series of autocorrelated states.  Estimates obtained by averaging over the time series of states generated from a given initial condition are referred to as time averaging. Figure~\ref{time_avrg_L} shows a comparison of the connectance obtained from time averaging and the solution of the mean field model~\eqref{thermodynamic_MF_equation_of_state} and the transition formula~\eqref{nushock} .

\begin{figure}[h]
\begin{center}
\includegraphics[width=8.4cm]{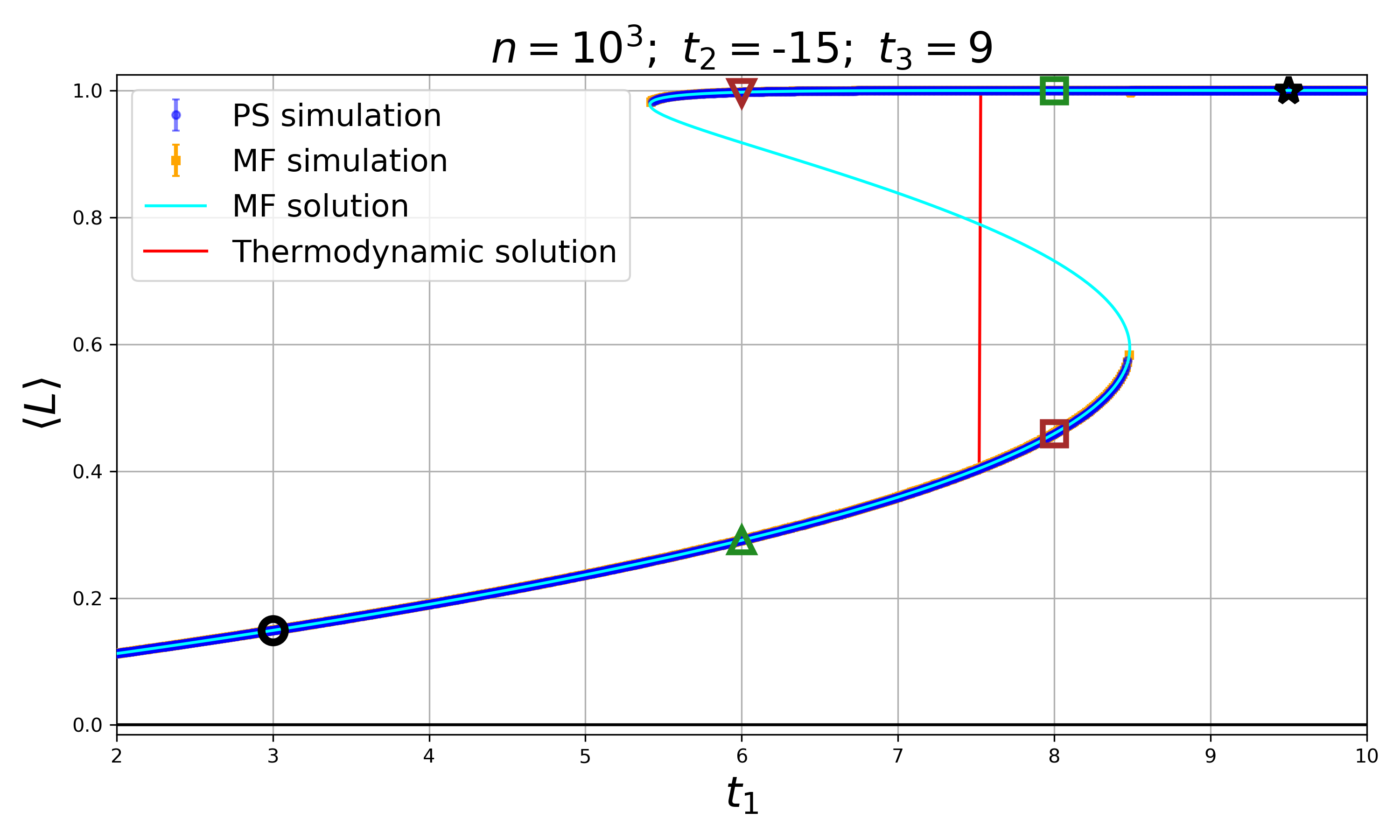} \\  
\includegraphics[width=4.2cm]{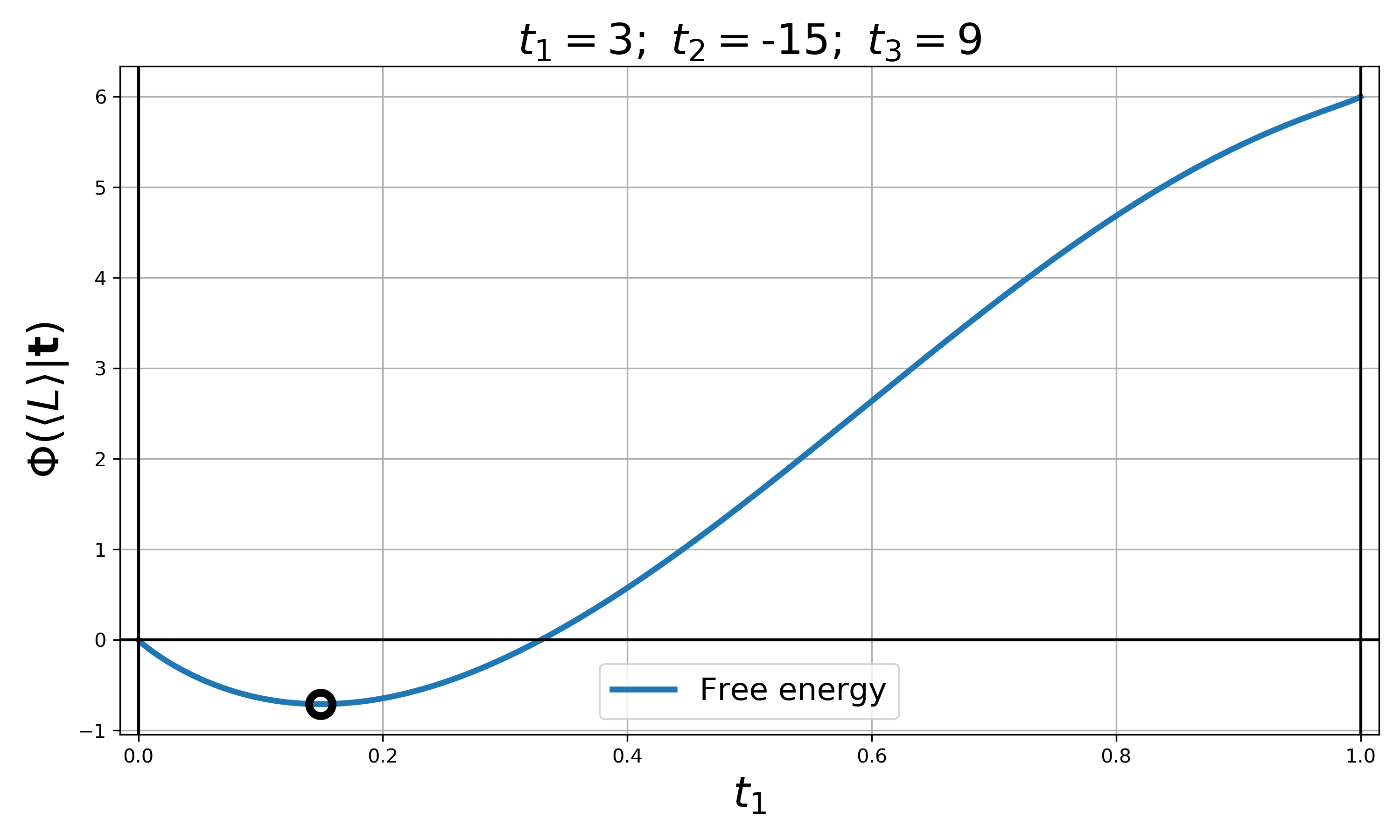}  \includegraphics[width=4.2cm]{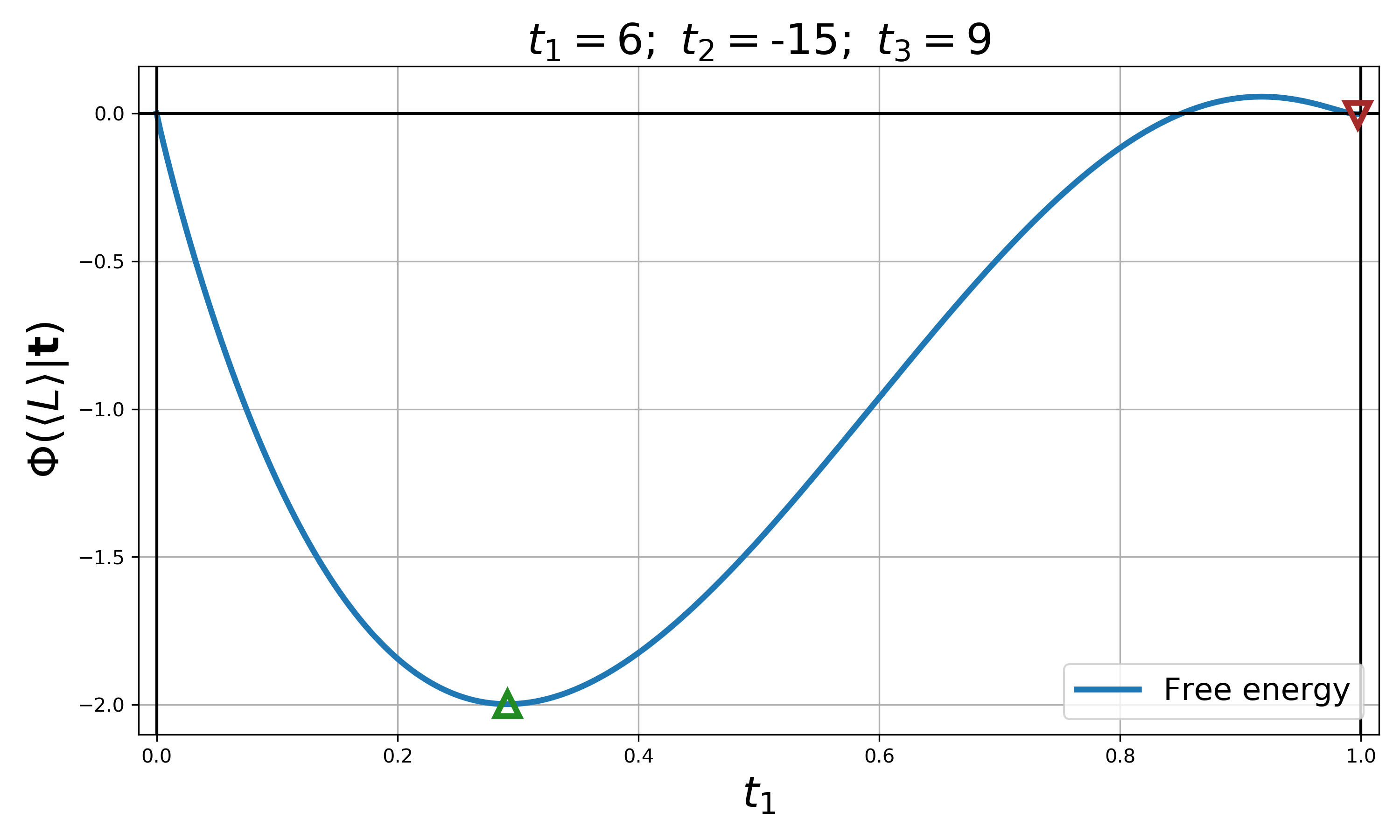}   \\\includegraphics[width=4.2cm]{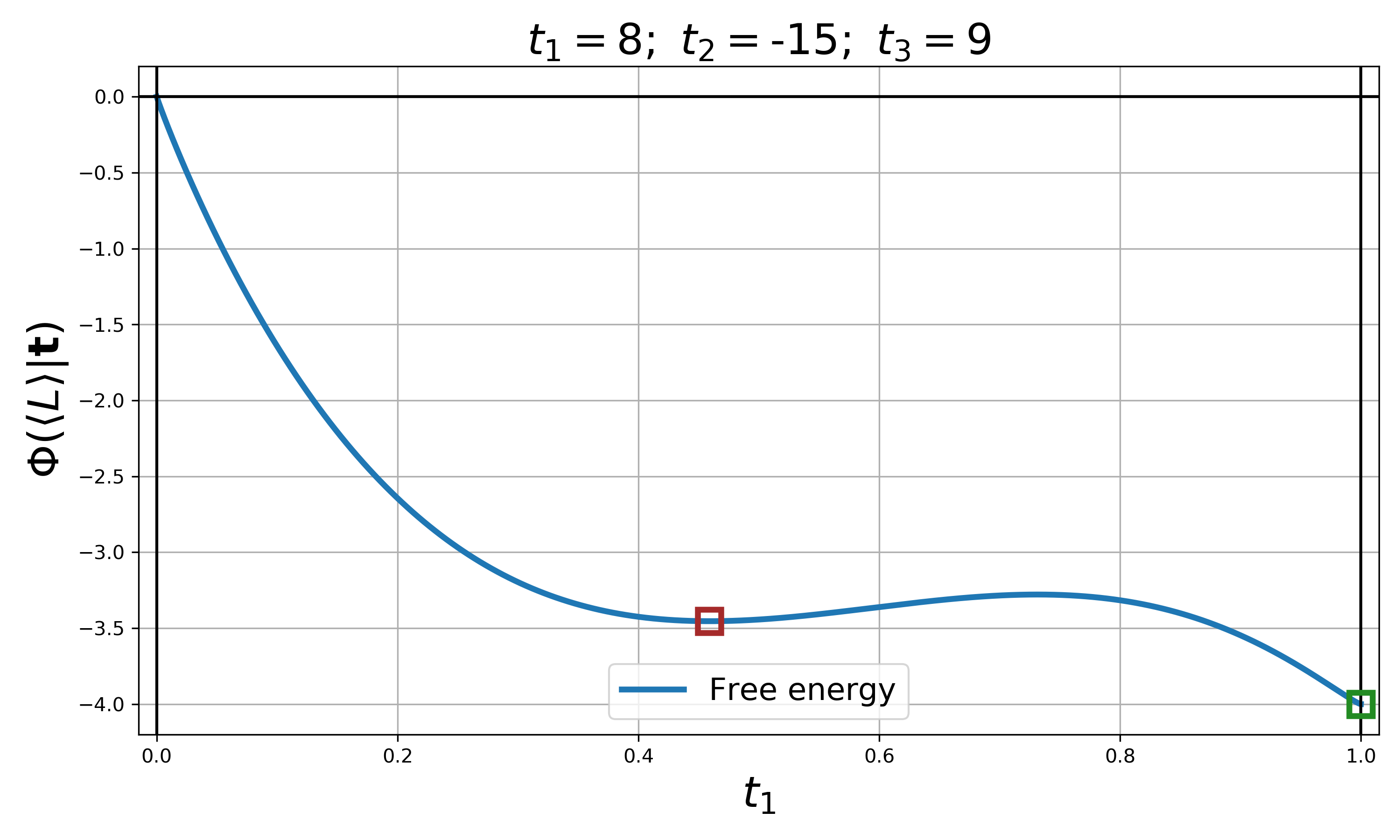}     \includegraphics[width=4.2cm]{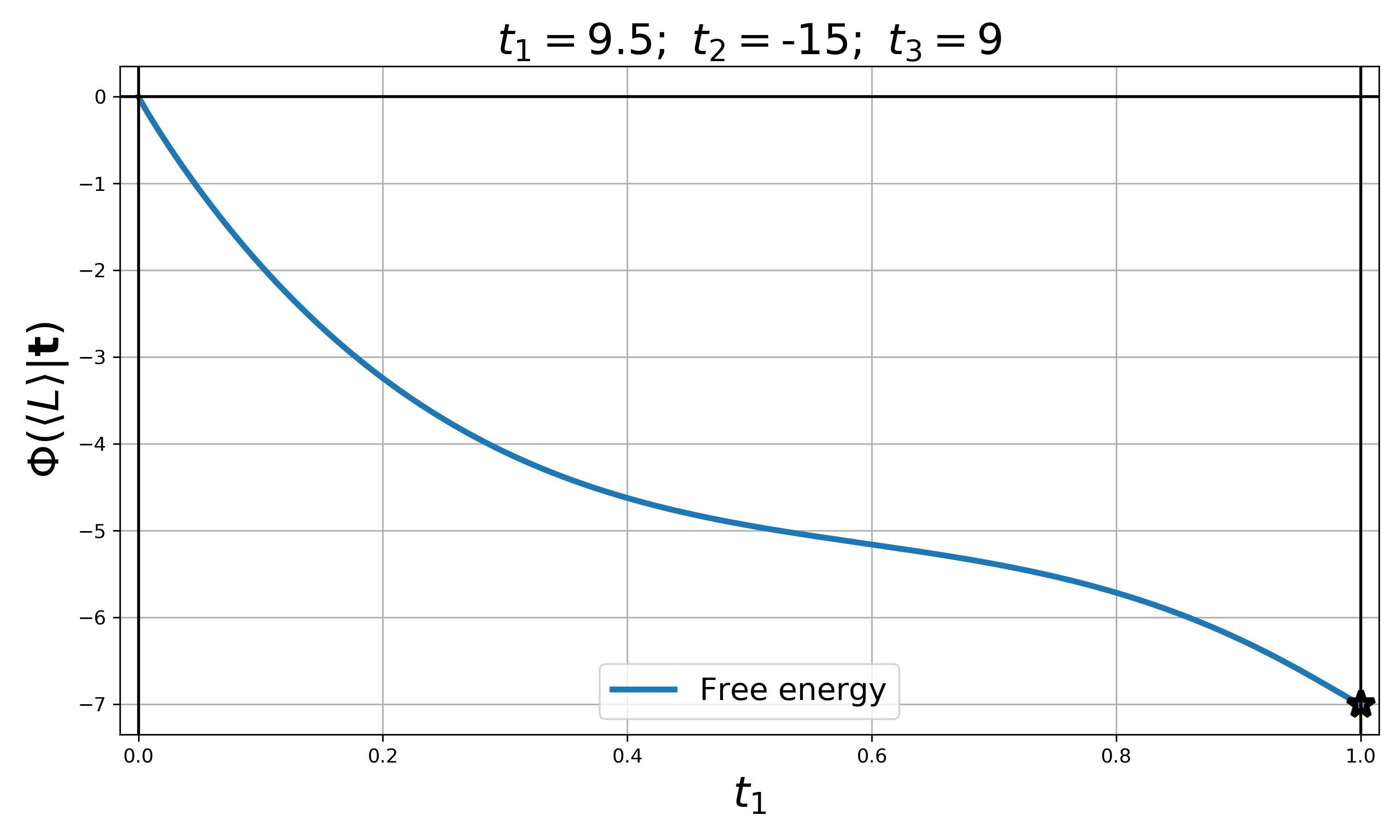}  \\
\end{center}
\caption{\small{Profile of the free energy for the $3-$star model and its MF analogue  for fixed values of $t_{1}$, $t_{2}$ and $t_{3}$ is compared against the connectance $\av{L}$ for the same values of $t_{2}$ and $t_{3}$ as $t_{1}$ changes. The lowest minimum (green cross) describes the stable branch of the solution $\av{L}$. Meta-stable states are associated to local minima (red cross) and correspond to the additional valued of $\av{L}$ in the region of multivaluedness. The red line superimposed to the connectance profile is obtained from direct numerical integration of the heat hierarchy and identifies the position of the shock connecting two stable branches.}}
\label{FE_Connectance}
\end{figure}

For sufficiently small networks, it is possible to perform a sufficient number of iterations such that the autocorrelation decays even in regimes where the system admits meta-stable states. This allows us to obtain accurate results for the averages over the actual canonical ensembles. Estimates of observables obtained by averaging over the realisations of the system starting from random initialisation are called ensemble averages. Ensemble averages capture stable states of the system and provide results that are consistent with the solution of the Burgers' hierarchy (\ref{burgersbruno}).

In this section we analyse the analog of the Helmholtz free energy $\Phi$ for the MF model, defined as, 
\begin{equation}
\label{phiMF}
\Phi = - 2 F = -2  \sum_{k=1}^p \langle L \rangle^kt_k + \log\left[\langle L\rangle^{\langle L\rangle }\left(1-\langle L\rangle\right)^{1-\langle L\rangle} \right],
\end{equation}
where $\av{L}$ is a solution of Eq. \eqref{thermodynamic_MF_equation_of_state}. We then compare the predicted observables with the Monte Carlo simulations for the $p$-star model and its mean field approximation. For illustrative purposes we consider the case $p=3$.

We should point out that the free energy \eqref{F_MF}  (or equivalently \eqref{phiMF}), expressed in terms of the solutions \eqref{thermodynamic_MF_equation_of_state} of the Hopf hierarchy, is derived under the assumption that $\av{L}$ is continuous and differentiable. This assumption is fulfilled in the region of the space of coupling constants $t_{k}$ such that the solution is single valued, but cannot be continuously extended across the singular sector (as illustrated for example in Figure \ref{fig:critA}), to the region where the solution is multivalued. 
However, in the region where the solution is multivalued, the expression \eqref{thermodynamic_MF_equation_of_state} allows to calculate the free energy of the system by selecting the branch where the free energy attains a minimum.

For instance, Figure~\ref{FE_standalone} illustrates the free energy profile as a function of $\av{L}$ for a choice of coupling constants such that $\av{L}$ is multivalued. The free energy $\Phi\left(\langle L\rangle|\mathbf t\right)$ admits two local minima of different depth in the close vicinity of $0$ and $1$. Based on the classical thermodynamic description, the absolute minimum corresponds to the stable state and the relative minimum is associated to a meta-stable state. 
MC simulations show that the GB distribution (\ref{GB_distribution}) realised via a  Markovian dynamics is such that, if the system is initialised in a meta-stable state, the system  will remain in that state for exponentially long times as a function of the number of links $N$ \cite{Fuller1963}. Therefore, for any finite time scale and for sufficiently large $N$, a system can be considered as non-ergodic, which implies that averages over time differ from averages over ensembles.

In Figure~\ref{time_avrg_L} we compare the connectance for the MF and $p$-star models in the thermodynamic regime. The comparison shows that the for $n = 10^{3}$ the system is in the thermodynamic regime. The analytic solution \eqref{thermodynamic_MF_equation_of_state} for the MF models in the thermodynamic limit is consistent with time averages of MC simulations for both models. In particular, the multivalued solution of the equation \eqref{thermodynamic_MF_equation_of_state} captures the meta-stable states  observed in MC simulations via suitable initialisation of the system. The transition formula \eqref{nushock} accurately captures the stable states for both models in the thermodynamic regime.
 
Figure~\ref{FE_Connectance} shows the profile of the free energy $\Phi\left(\langle L\rangle|\mathbf t\right)$ for the $3$-star model and its MF analogue  for fixed values of $t_{1}$, $t_{2}$ and $t_{3}$, compared against the connectance  $\av{L}$ as a function of $t_{1}$ for the same values of $t_{2}$ and $t_{3}$. The lowest minimum  describes the stable branch of the solution $\av{L}$. Meta-stable states are associated to local minima and correspond to the additional valued of $\av{L}$ in the region of multivaluedness. 

The position, the depth and number of local minima of $\Phi\left(\langle L\rangle|\mathbf t\right)$ change with $t_{k}$ and, when a local minimum turns into a global one as $t_{1}$ increases, ensemble averages of the observables develop a jump. The branches of the curve (\ref{thermodynamic_MF_equation_of_state}) on which $\langle L\rangle$ decreases with $t_1$ correspond to the (local) maxima of $\Phi\left(\langle L\rangle|\mathbf t\right)$ and therefore are not stable.

\begin{figure}[h]
  \begin{center}
    \includegraphics[width=7.5cm]{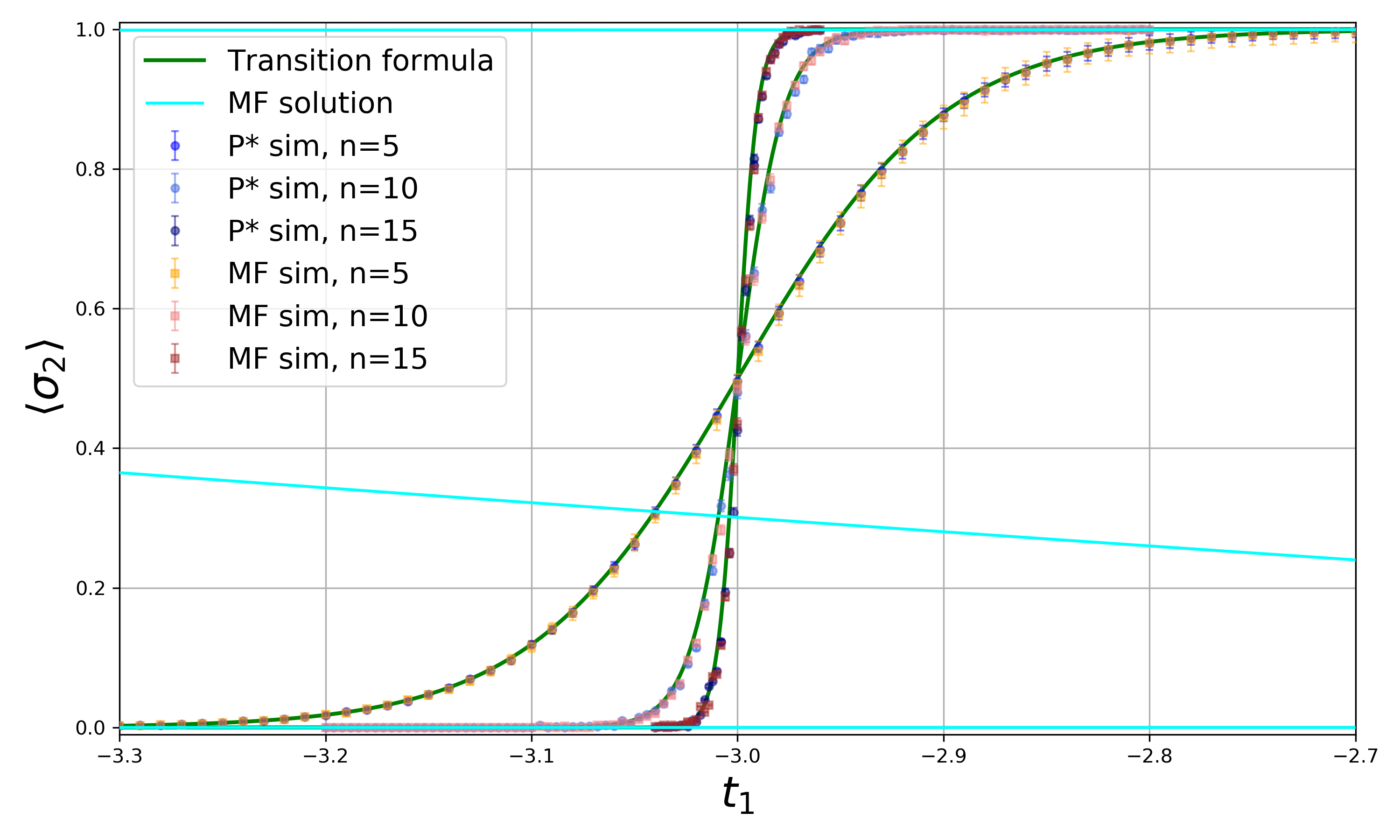} \\    \includegraphics[width=7.5cm]{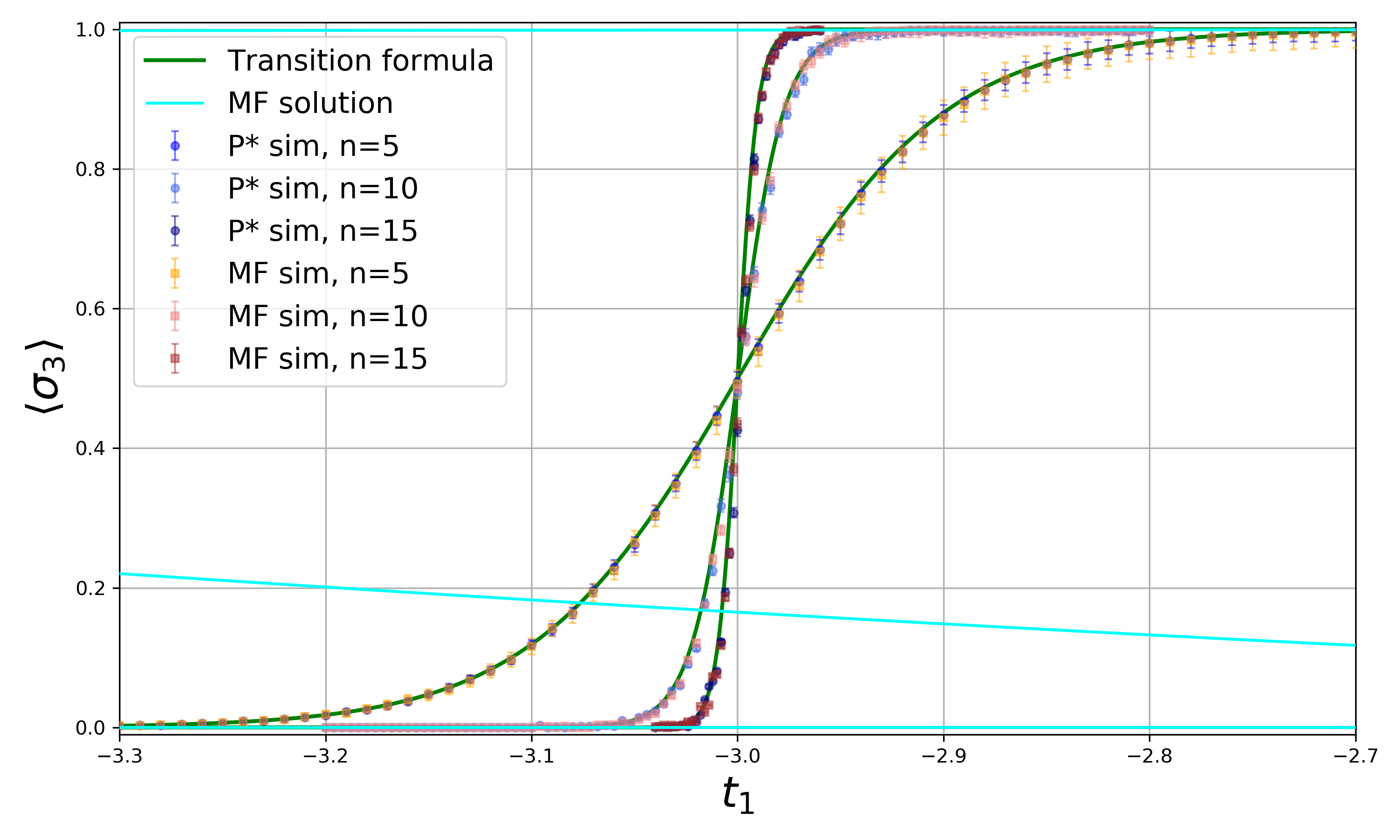}
  \end{center}
  \caption{\small{Comparison of simulated densities of $2-$stars and $3-$stars for the $3-$star model and moments $\av{L^{2}}$ and $\av{L^{3}}$ for the corresponding MF model for different sizes $n$. For illustrative purposes we chose  $t_{2} = 2$, $t_{3} = 1$. Multivalued solution for the MF model in the thermodynamic limit is also shown for reference.  In this regime, MF model and transition formula \eqref{nushock} provide an accurate description of $p$-star model.}}
  \label{stars_ensemble_averages}
\end{figure}

In Figure~\ref{stars_ensemble_averages} we compare the expectation values, via ensemble averages, of the density of $k$-stars in the $p$-star model, defined as
\begin{equation} \label{sigmas}
  \sigma_k \equiv \frac{1}{n}\binom{n-1}{k}^{-1}S_k\in[0,1],
\end{equation}
and the moments $\av{L^{k}}$ for the MF model. For illustrative purposes we choose again $p=3$. In the region of values for  the coupling constants where the free energy has a single local minimum, the hierarchy of equations~\eqref{HJeq} holds in the thermodynamic regime, and implies the equality
\begin{equation} 
\av{L^{k}} = \av{L}^{k} \qquad k=1,2,3,\dots
\end{equation}
In this regime, the $k$-star densities, estimated via MC simulations, are well approximated by the corresponding $k$-th moments of the  MF model, for $k=2$ and $k=3$. This result is consistent with existing studies on the mean field theory approximation in the thermodynamic limit, see e.g., \cite{ParkNewman2005}. Furthermore,  Figure~\ref{stars_ensemble_averages}, shows that, even for small networks, the mean field theory is accurate sufficiently far from the transition region. In the case where the system admits $(0,1)$-bistability, formula (\ref{semiempirical}) provides ensemble averages of the observables in the transition region. 

\section{Local properties: MF vs $p$-star model} 
\label{distributions_subsec}

It is interesting to compare the MF and $p$-star models relative to some local features  such as the expectations $\av{k_{i}}$ of  the node degree,
\begin{equation*}
\label{nodeg}
k_{i} = \sum_{j=1}^{n} A_{ij},
\end{equation*}
which give the number links attached to a given node, and the expectation $\av{c_{i}}$ of the local clustering coefficient (LCC)
\begin{equation} \label{ER_lcc}
c_i(\mathbf A) = \frac{\sum_{j,k}A_{ij}A_{jk}A_{ki}}{\sum_{j,l}A_{ij}A_{il}\left(1-\delta_{jl}\right)},
\end{equation}
which gives the ratio between triangles and $2-$star in the network.
For illustrative purposes, the MC simulations are performed for the case $p=3$.
Hence, we write the Hamiltonian of the $3$-star model in terms of the node degree as follows
\begin{equation}\label{HPS_2}
  H(\mathbf A) = -\epsilon_1\sum_{i=1}^nk_i - \frac{\epsilon_2}{2n}\sum_{i=1}^nk_i^2 - \frac{\epsilon_3}{6n^2}\sum_{i=1}^nk_i^3,
\end{equation}
where
\begin{multline}
  \epsilon_1 = \tau_1-\frac{\tau_2}{2n}+\frac{\tau_3}{3n^2}\qquad \epsilon_2=\tau_2-\frac{\tau_3}{n} \qquad
  \epsilon_3 = \tau_3.
\end{multline}
As observed in \cite{FrankStrauss1986}, the Hamiltonian of the general homogeneous ERGM with interactions among links sharing a node can be written as sum of the Hamiltonian (\ref{Hstar}) and a term proportional to the number of triangles in the graph. It is therefore interesting to study how the LCC behaves in the $p$-star model {\it per se}, in absence of extra contribution from triangles.
\begin{figure}[h]
\begin{center}
\includegraphics[width=7.5cm]{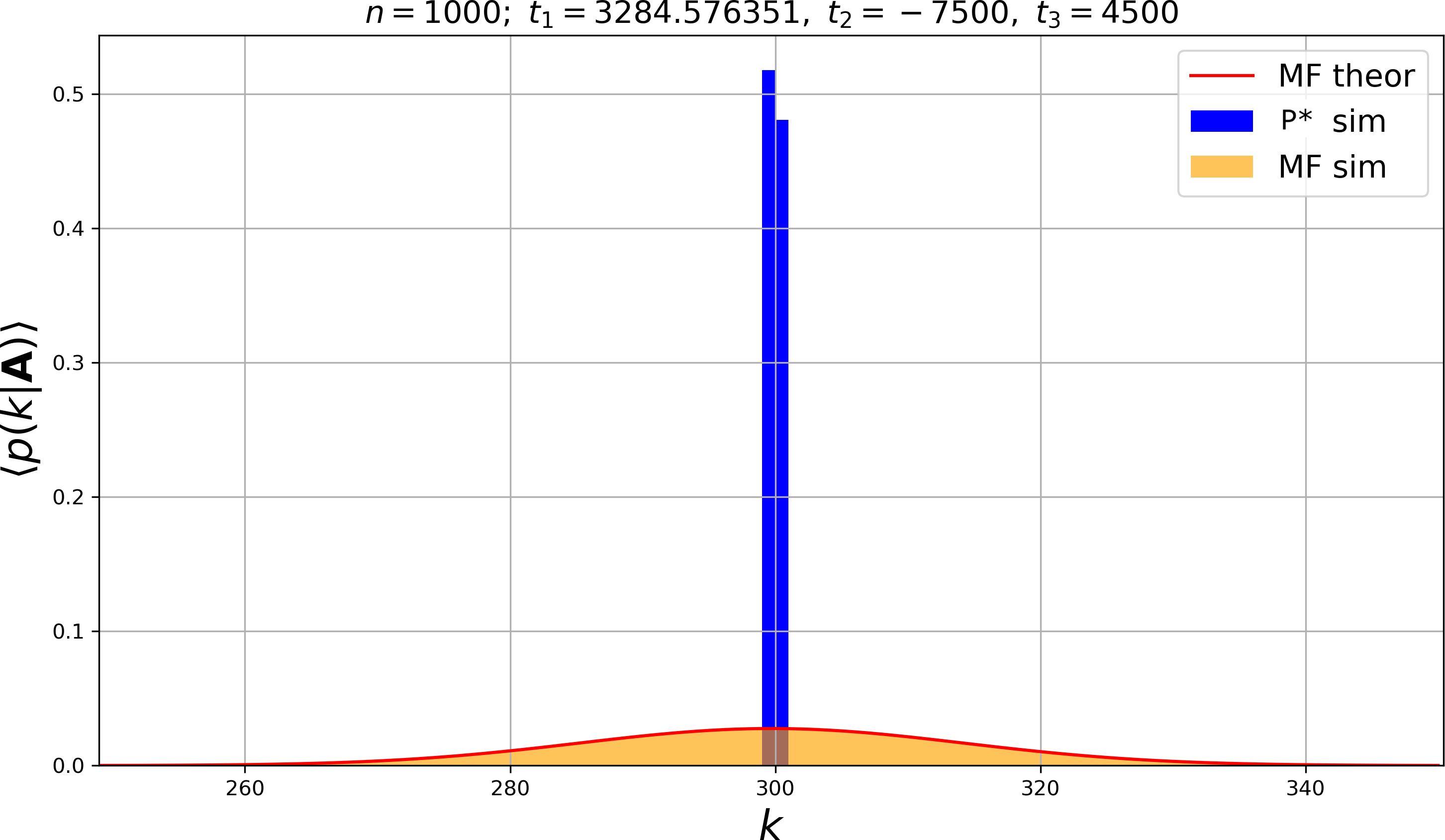} \\ \includegraphics[width=7.5cm]{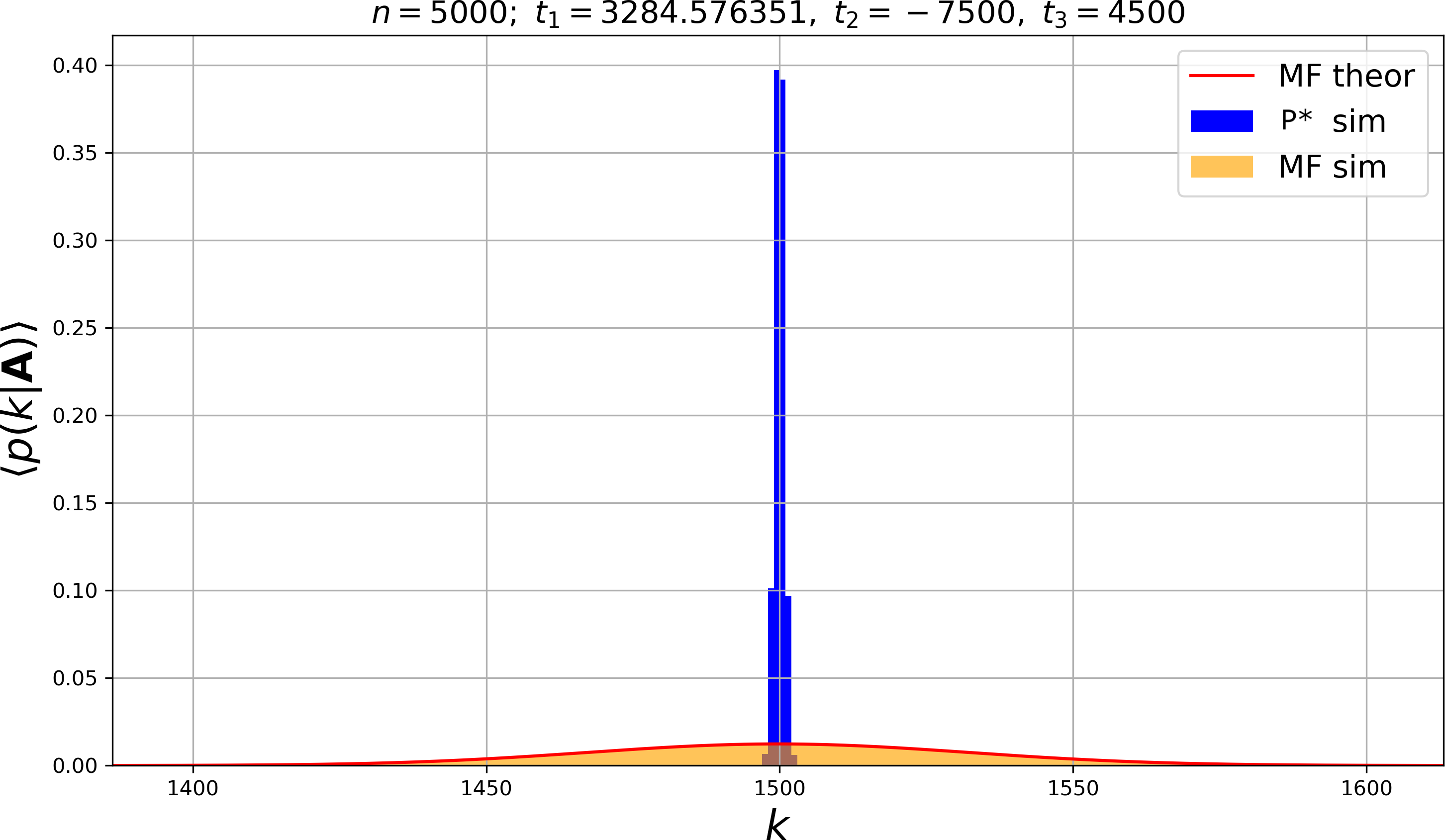} 
\end{center}
\caption{\small{Degree distributions of  $p$-star and MF models with $1000$ and $5000$ nodes at for large values $|t_{k}|$, with $k=1,2,3$. The degree distribution of the $p$-star is almost degenerate around dominant degrees $k \sim 300$. Degree distribution for the MF model follows instead the expression (\ref{approx_ER_deg_distr}).}}
\label{low_temp_distribs2}
\end{figure}

\begin{figure}[h]
\begin{center}
\includegraphics[width=7.5cm]{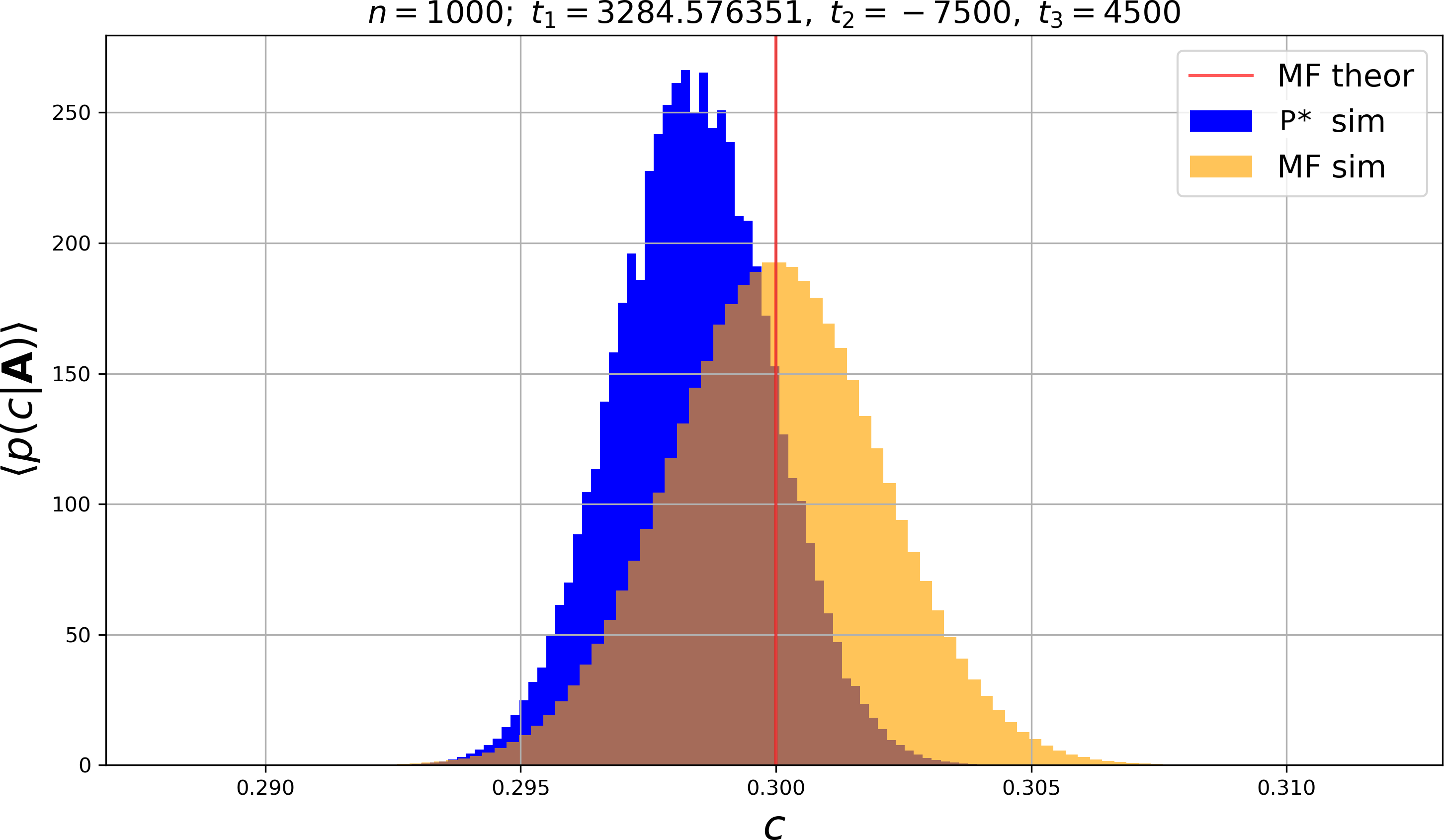} \\ \includegraphics[width=7.5cm]{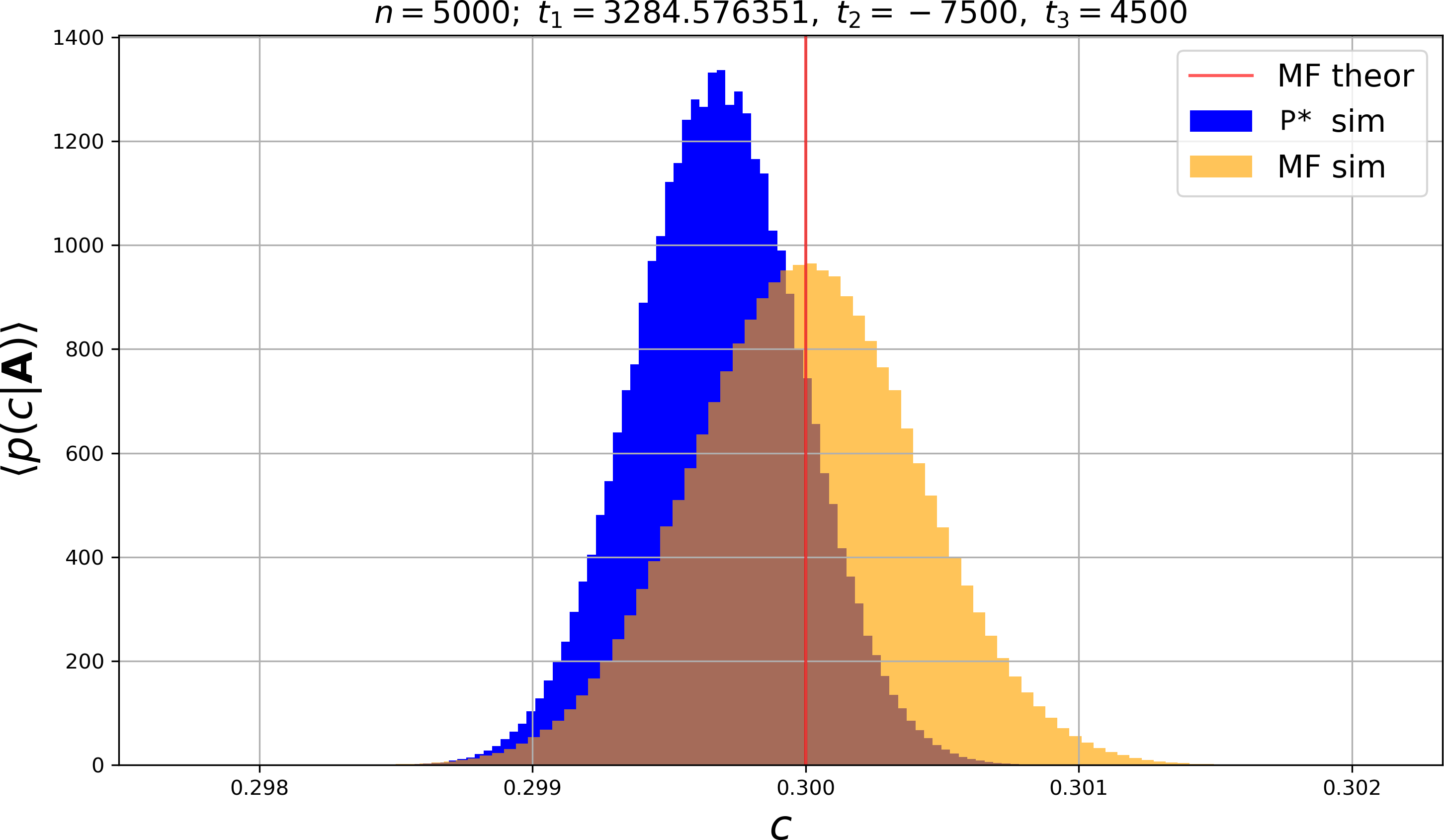}  
\end{center}
\caption{\small{LCC distributions of  $p$-star and MF models with $1000$ and $5000$ nodes for large values $|t_{k}|$ with $k=1,2,3$.}}
\label{low_temp_distribs}
\end{figure}

We have that the degree distribution of the MF model follows a binomial distribution (as the ER model),  i.e.,
\begin{equation}
  \label{ER_deg_distr}
  \langle p(k|\mathbf A)\rangle = \binom{n-1}{k}L^k(1-L)^{N-1-k},
\end{equation}
which, for large graphs, and with moderate connectance $p(k|\mathbf A)\rangle$, is approximated by the Gaussian distribution
\begin{equation}
  \label{approx_ER_deg_distr}
  \langle p(k|\mathbf A)\rangle \simeq \frac{1}{\sqrt{2\pi nL(1-L)}}\mathrm{e}^{-\frac{(k-Ln)^2}{2nL(1-L)}}.
\end{equation}
This is a consequence of the fact that the MF Hamiltonian (\ref{HamL}) depends solely on the connectance, and therefore all possible graphs with the same total number of links occur with the same probability. 
We also observe that  the $p$-star model does not have this kind of degeneracy and, at low temperatures, produces mostly regular graphs, as shown in Figure~\ref{low_temp_distribs2}. The LCC distributions are qualitatively similar for both the $p$-star and the MF model, as shown in Figure~\ref{low_temp_distribs} although, as expected, for the MF model triangles tend to be more dominant over $2-$star compared to the $p$-star model.

\begin{figure}[h]
\begin{center}
\includegraphics[width=8.5cm]{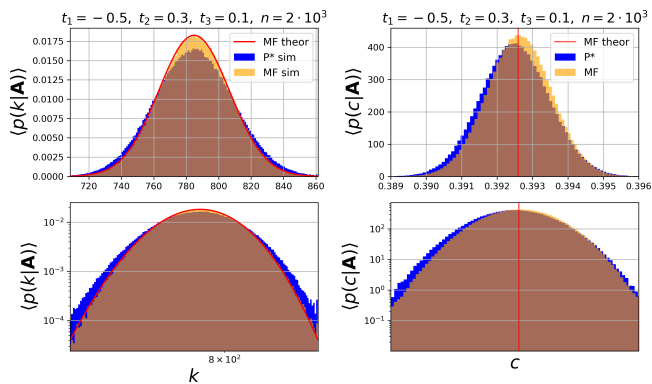}
\end{center}
\caption{\small{Degree distributions and local clustering coefficients for $p$-star and MF models at $t_1=-0.5,\ t_2=0.3,\ t_3=0.1$ ({\it top row}) and corresponding log-log plot ({\it bottom row}) . The degree distribution of the MF model is described well by the Gaussian distribution (\ref{approx_ER_deg_distr}) (red curve on degree distribution plots). The vertical red lines on LCC plots show the position of the average LCC given by (\ref{ER_lcc}). Both degree and LCC distributions of the 3-star model in this case are wider than the degree distribution of the corresponding MF model but the averages match well. Decay is exponential in both cases as shown in log-log plots.}}
\label{deg_dist}
\end{figure}

Figure~\ref{deg_dist} shows that the degree distributions of the $p$-star model may also be wider than predicted by (\ref{ER_deg_distr}). This fact can be explained by the tendency of the $p$-star model with positive couplings to create hubs \cite{AnnibaleCourtney2015}, i.e., nodes with high degrees and, therefore, high numbers of stars. However, the decay of the distributions appears to be exponential for both models,  as shown by the log-log plots of Figure~\ref{deg_dist} (bottom). 

We finally note that while the functional form of the LCC distributions of the $p$-star and MF models  shown in Figure~\ref{deg_dist} and Figure~\ref{low_temp_distribs} is different, both distributions get narrower as $n$ increases.
As for the degree distribution, depending on parameters, the LCC distributions of the $p$-star model may be slightly wider (Figure~\ref{deg_dist}) or narrower (Figure~\ref{low_temp_distribs}) than that of the MF model.  However, the tails of the distributions are exponential for both models.

\section{Concluding remarks}
\label{sec:conclusions}
A detailed comparison between $p$-star models defined by the Hamiltonian \eqref{Hstar} and their mean field analogues defined by the Hamitonian \eqref{Hamp} shows that, in the thermodynamic regime, the latter captures with high accuracy  both macroscopic qualitative and quantitative features of the former. However, we have observed that discrepancies arise at the microscopic level when local properties, such as local clustering coefficients, are compared. This is indeed not surprising, as the Hamiltonian of the mean field analogue admits an explicit expression in terms of global variables such as the connectance and therefore it is not devised to discern details of local properties.

Mean field models, i.e., models defined on a fully connected graph with arbitrary degree of interaction, are interesting in both the finite size and the thermodynamic regimes for their formal as well as phenomenological properties. Indeed, the MF model is a completely integrable system, as the partition function satisfies a compatible system of linear partial differential equations (differential identities) represented by the heat hierarchy and the solution is specified by an initial condition corresponding to the Hamiltonian \eqref{Hamp} evaluated at zero coupling constants. Therefore, once the suitable differential identities for the partition function,  corresponding to the heat hierarchy, are given, the solution of the model is reduced to the solution of the ER model. The heat hierarchy leads to an explicit formula for both the finite size and the thermodynamic regime. The thermodynamic limit is calculated based on the scaling properties of the heat hierarchy, and the free energy satisfies a hierarchy Hamilton-Jacobi type equations whose differential consequences allow to calculate corresponding equations for the state functions and the order parameters. In the thermodynamic limit, explicit equations of state are obtained by solving the hierarchy via the method of characteristics. 

We also observed that there is a one-to-one correspondence between the thermodynamic solutions of the MF models and the approximation of the $p$-star models obtained from the minimisation of Kulbach-Leibler divergence. The MC simulations confirm the qualitative and quantitative agreement of the two models in the specified regime.
Hence, the asymptotic study of the thermodynamic system via the heat hierarchy provides an effective approach to the description of finite size effects leading to the resolution of singularity in the thermodynamic limit as well as the analytic description of order parameters in the transition region. In comparison, obtaining the same results via direct Metropolis-Hastings algorithm requires extensive simulations due to slow convergence induced by the presence of meta-stable states. We finally note that the approach based on the use of differential identities described above applies to a variety of systems, from classical magnetic systems \cite{DeNittisMoro2012,MoroAnnals2014,ABDM2016,BarraMoro2014,GiglioDeMatteisMoro2018,LorenzoniMoro2019,GenoveseBarra2009}  to random matrix models \cite{BenassiMoro2020,BenassiDellAttiMoro2021} and it proves to be effective for the analytic description of systems of increasing complexity.

{\it Acknowledgements}. This research is supported by Royal Society International Exchanges grant 
IES$ \backslash $ R2$ \backslash $170116, the Leverhulme Trust Research Project grant RPG 2017-228 and the National Science Foundation under Grant No. DMS-2009487.

\bibliography{biblio.bib}

\begin{thebibliography}{33}%
\makeatletter
\providecommand \@ifxundefined [1]{%
 \@ifx{#1\undefined}
}%
\providecommand \@ifnum [1]{%
 \ifnum #1\expandafter \@firstoftwo
 \else \expandafter \@secondoftwo
 \fi
}%
\providecommand \@ifx [1]{%
 \ifx #1\expandafter \@firstoftwo
 \else \expandafter \@secondoftwo
 \fi
}%
\providecommand \natexlab [1]{#1}%
\providecommand \enquote  [1]{``#1''}%
\providecommand \bibnamefont  [1]{#1}%
\providecommand \bibfnamefont [1]{#1}%
\providecommand \citenamefont [1]{#1}%
\providecommand \href@noop [0]{\@secondoftwo}%
\providecommand \href [0]{\begingroup \@sanitize@url \@href}%
\providecommand \@href[1]{\@@startlink{#1}\@@href}%
\providecommand \@@href[1]{\endgroup#1\@@endlink}%
\providecommand \@sanitize@url [0]{\catcode `\\12\catcode `\$12\catcode
  `\&12\catcode `\#12\catcode `\^12\catcode `\_12\catcode `\%12\relax}%
\providecommand \@@startlink[1]{}%
\providecommand \@@endlink[0]{}%
\providecommand \url  [0]{\begingroup\@sanitize@url \@url }%
\providecommand \@url [1]{\endgroup\@href {#1}{\urlprefix }}%
\providecommand \urlprefix  [0]{URL }%
\providecommand \Eprint [0]{\href }%
\providecommand \doibase [0]{https://doi.org/}%
\providecommand \selectlanguage [0]{\@gobble}%
\providecommand \bibinfo  [0]{\@secondoftwo}%
\providecommand \bibfield  [0]{\@secondoftwo}%
\providecommand \translation [1]{[#1]}%
\providecommand \BibitemOpen [0]{}%
\providecommand \bibitemStop [0]{}%
\providecommand \bibitemNoStop [0]{.\EOS\space}%
\providecommand \EOS [0]{\spacefactor3000\relax}%
\providecommand \BibitemShut  [1]{\csname bibitem#1\endcsname}%
\let\auto@bib@innerbib\@empty
\bibitem [{\citenamefont {Newman}(2010)}]{Newman_book}%
  \BibitemOpen
  \bibfield  {author} {\bibinfo {author} {\bibfnamefont {M.~E.~J.}\
  \bibnamefont {Newman}},\ }\href@noop {} {\emph {\bibinfo {title} {{Networks:
  an Introduction}}}}\ (\bibinfo  {publisher} {{Oxford University Press, New
  York}},\ \bibinfo {year} {2010})\BibitemShut {NoStop}%
\bibitem [{\citenamefont {Barab\'asi}(2016)}]{Barabasi_book}%
  \BibitemOpen
  \bibfield  {author} {\bibinfo {author} {\bibfnamefont {A.-L.}\ \bibnamefont
  {Barab\'asi}},\ }\href@noop {} {\emph {\bibinfo {title} {Network Science}}}\
  (\bibinfo  {publisher} {Cambridge University Press},\ \bibinfo {year}
  {2016})\BibitemShut {NoStop}%
\bibitem [{\citenamefont {Barab\'asi}\ and\ \citenamefont
  {Oltvai}(2004)}]{Barabasi_Network_Biology}%
  \BibitemOpen
  \bibfield  {author} {\bibinfo {author} {\bibfnamefont {A.-L.}\ \bibnamefont
  {Barab\'asi}}\ and\ \bibinfo {author} {\bibfnamefont {Z.}~\bibnamefont
  {Oltvai}},\ }\bibfield  {title} {\bibinfo {title} {Network biology:
  understanding the cell's functional organization},\ }\href@noop {} {\bibfield
   {journal} {\bibinfo  {journal} {Nature Reviews Genetics}\ }\textbf {\bibinfo
  {volume} {5}},\ \bibinfo {pages} {101} (\bibinfo {year} {2004})}\BibitemShut
  {NoStop}%
\bibitem [{\citenamefont {Stein}\ \emph {et~al.}(2015)\citenamefont {Stein},
  \citenamefont {Marks},\ and\ \citenamefont {Sander}}]{Stein_Marks_Sander}%
  \BibitemOpen
  \bibfield  {author} {\bibinfo {author} {\bibfnamefont {R.}~\bibnamefont
  {Stein}}, \bibinfo {author} {\bibfnamefont {D.~S.}\ \bibnamefont {Marks}},\
  and\ \bibinfo {author} {\bibfnamefont {C.}~\bibnamefont {Sander}},\
  }\bibfield  {title} {\bibinfo {title} {Inferring pairwise interactions from
  biological data using maximum-entropy probability models},\ }\href@noop {}
  {\bibfield  {journal} {\bibinfo  {journal} {PLoS Comput. Biol.}\ }\textbf
  {\bibinfo {volume} {11}},\ \bibinfo {pages} {e1004182} (\bibinfo {year}
  {2015})}\BibitemShut {NoStop}%
\bibitem [{\citenamefont {Bassett}\ and\ \citenamefont
  {Sporns}(2017)}]{Bassett2017}%
  \BibitemOpen
  \bibfield  {author} {\bibinfo {author} {\bibfnamefont {D.}~\bibnamefont
  {Bassett}}\ and\ \bibinfo {author} {\bibfnamefont {O.}~\bibnamefont
  {Sporns}},\ }\bibfield  {title} {\bibinfo {title} {Network neuroscience},\
  }\href@noop {} {\bibfield  {journal} {\bibinfo  {journal} {Nat. Neurosci.}\
  }\textbf {\bibinfo {volume} {20}},\ \bibinfo {pages} {353} (\bibinfo {year}
  {2017})}\BibitemShut {NoStop}%
\bibitem [{\citenamefont {Robins}\ \emph
  {et~al.}(2007{\natexlab{a}})\citenamefont {Robins}, \citenamefont {Pattison},
  \citenamefont {Kalish},\ and\ \citenamefont {Lusher}}]{ROBINS2007173}%
  \BibitemOpen
  \bibfield  {author} {\bibinfo {author} {\bibfnamefont {G.}~\bibnamefont
  {Robins}}, \bibinfo {author} {\bibfnamefont {P.}~\bibnamefont {Pattison}},
  \bibinfo {author} {\bibfnamefont {Y.}~\bibnamefont {Kalish}},\ and\ \bibinfo
  {author} {\bibfnamefont {D.}~\bibnamefont {Lusher}},\ }\bibfield  {title}
  {\bibinfo {title} {An introduction to exponential random graph (p*) models
  for social networks},\ }\href@noop {} {\bibfield  {journal} {\bibinfo
  {journal} {Social Networks}\ }\textbf {\bibinfo {volume} {29}},\ \bibinfo
  {pages} {173} (\bibinfo {year} {2007}{\natexlab{a}})}\BibitemShut {NoStop}%
\bibitem [{\citenamefont {Robins}\ \emph
  {et~al.}(2007{\natexlab{b}})\citenamefont {Robins}, \citenamefont {Snijders},
  \citenamefont {Wang}, \citenamefont {Handcock},\ and\ \citenamefont
  {Pattison}}]{ROBINS2007192}%
  \BibitemOpen
  \bibfield  {author} {\bibinfo {author} {\bibfnamefont {G.}~\bibnamefont
  {Robins}}, \bibinfo {author} {\bibfnamefont {T.}~\bibnamefont {Snijders}},
  \bibinfo {author} {\bibfnamefont {P.}~\bibnamefont {Wang}}, \bibinfo {author}
  {\bibfnamefont {M.}~\bibnamefont {Handcock}},\ and\ \bibinfo {author}
  {\bibfnamefont {P.}~\bibnamefont {Pattison}},\ }\bibfield  {title} {\bibinfo
  {title} {Recent developments in exponential random graph (p*) models for
  social networks},\ }\href@noop {} {\bibfield  {journal} {\bibinfo  {journal}
  {Social Networks}\ }\textbf {\bibinfo {volume} {29}},\ \bibinfo {pages} {192}
  (\bibinfo {year} {2007}{\natexlab{b}})}\BibitemShut {NoStop}%
\bibitem [{\citenamefont {Anderson}\ \emph {et~al.}(1999)\citenamefont
  {Anderson}, \citenamefont {Wasserman},\ and\ \citenamefont
  {Crouch}}]{ANDERSON199937}%
  \BibitemOpen
  \bibfield  {author} {\bibinfo {author} {\bibfnamefont {C.~J.}\ \bibnamefont
  {Anderson}}, \bibinfo {author} {\bibfnamefont {S.}~\bibnamefont
  {Wasserman}},\ and\ \bibinfo {author} {\bibfnamefont {B.}~\bibnamefont
  {Crouch}},\ }\bibfield  {title} {\bibinfo {title} {A p* primer: logit models
  for social networks},\ }\href@noop {} {\bibfield  {journal} {\bibinfo
  {journal} {Social Networks}\ }\textbf {\bibinfo {volume} {21}},\ \bibinfo
  {pages} {37} (\bibinfo {year} {1999})}\BibitemShut {NoStop}%
\bibitem [{\citenamefont {Wasserman}\ and\ \citenamefont
  {Pattison}(1996)}]{WassPatt1996}%
  \BibitemOpen
  \bibfield  {author} {\bibinfo {author} {\bibfnamefont {S.}~\bibnamefont
  {Wasserman}}\ and\ \bibinfo {author} {\bibfnamefont {P.}~\bibnamefont
  {Pattison}},\ }\bibfield  {title} {\bibinfo {title} {Logit models and
  logistic regressions for social networks: I. {A}n introduction to {M}arkov
  graphs and $p*$},\ }\href@noop {} {\bibfield  {journal} {\bibinfo  {journal}
  {Psychometrika}\ }\textbf {\bibinfo {volume} {61}},\ \bibinfo {pages} {401}
  (\bibinfo {year} {1996})}\BibitemShut {NoStop}%
\bibitem [{\citenamefont {Pastor-Satorras}\ \emph {et~al.}(2015)\citenamefont
  {Pastor-Satorras}, \citenamefont {Castellano}, \citenamefont {Mieghem},\ and\
  \citenamefont {Vespignani}}]{Pastor2015}%
  \BibitemOpen
  \bibfield  {author} {\bibinfo {author} {\bibfnamefont {R.}~\bibnamefont
  {Pastor-Satorras}}, \bibinfo {author} {\bibfnamefont {C.}~\bibnamefont
  {Castellano}}, \bibinfo {author} {\bibfnamefont {P.~V.}\ \bibnamefont
  {Mieghem}},\ and\ \bibinfo {author} {\bibfnamefont {A.}~\bibnamefont
  {Vespignani}},\ }\bibfield  {title} {\bibinfo {title} {Epidemic processes in
  complex networks},\ }\href@noop {} {\bibfield  {journal} {\bibinfo  {journal}
  {Rev. Mod. Phys.}\ }\textbf {\bibinfo {volume} {87}},\ \bibinfo {pages} {925}
  (\bibinfo {year} {2015})}\BibitemShut {NoStop}%
\bibitem [{\citenamefont {Park}\ and\ \citenamefont
  {Newman}(2004)}]{ParkNewman2014}%
  \BibitemOpen
  \bibfield  {author} {\bibinfo {author} {\bibfnamefont {J.}~\bibnamefont
  {Park}}\ and\ \bibinfo {author} {\bibfnamefont {M.}~\bibnamefont {Newman}},\
  }\bibfield  {title} {\bibinfo {title} {Statistical mechanics of networks},\
  }\href@noop {} {\bibfield  {journal} {\bibinfo  {journal} {Phys. Rev. E}\
  }\textbf {\bibinfo {volume} {70}},\ \bibinfo {pages} {066117} (\bibinfo
  {year} {2004})}\BibitemShut {NoStop}%
\bibitem [{\citenamefont {Barra}\ \emph {et~al.}(2014)\citenamefont {Barra},
  \citenamefont {Lorenzo}, \citenamefont {Guerra},\ and\ \citenamefont
  {Moro}}]{BarraMoro2014}%
  \BibitemOpen
  \bibfield  {author} {\bibinfo {author} {\bibfnamefont {A.}~\bibnamefont
  {Barra}}, \bibinfo {author} {\bibfnamefont {A.~D.}\ \bibnamefont {Lorenzo}},
  \bibinfo {author} {\bibfnamefont {F.}~\bibnamefont {Guerra}},\ and\ \bibinfo
  {author} {\bibfnamefont {A.}~\bibnamefont {Moro}},\ }\bibfield  {title}
  {\bibinfo {title} {On quantum and relativistic mechanical analogues in
  mean-field spin models},\ }\href@noop {} {\bibfield  {journal} {\bibinfo
  {journal} {Proc. R. Soc. A}\ }\textbf {\bibinfo {volume} {470}},\ \bibinfo
  {pages} {20140589} (\bibinfo {year} {2014})}\BibitemShut {NoStop}%
\bibitem [{\citenamefont {Arsie}\ \emph
  {et~al.}(2015{\natexlab{a}})\citenamefont {Arsie}, \citenamefont
  {Lorenzoni},\ and\ \citenamefont {Moro}}]{ALM2015}%
  \BibitemOpen
  \bibfield  {author} {\bibinfo {author} {\bibfnamefont {A.}~\bibnamefont
  {Arsie}}, \bibinfo {author} {\bibfnamefont {P.}~\bibnamefont {Lorenzoni}},\
  and\ \bibinfo {author} {\bibfnamefont {A.}~\bibnamefont {Moro}},\ }\bibfield
  {title} {\bibinfo {title} {Integrable viscous conservation laws},\
  }\href@noop {} {\bibfield  {journal} {\bibinfo  {journal} {Nonlinearity}\
  }\textbf {\bibinfo {volume} {28}},\ \bibinfo {pages} {1859} (\bibinfo {year}
  {2015}{\natexlab{a}})}\BibitemShut {NoStop}%
\bibitem [{\citenamefont {Arsie}\ \emph
  {et~al.}(2015{\natexlab{b}})\citenamefont {Arsie}, \citenamefont
  {Lorenzoni},\ and\ \citenamefont {Moro}}]{ALM2015b}%
  \BibitemOpen
  \bibfield  {author} {\bibinfo {author} {\bibfnamefont {A.}~\bibnamefont
  {Arsie}}, \bibinfo {author} {\bibfnamefont {P.}~\bibnamefont {Lorenzoni}},\
  and\ \bibinfo {author} {\bibfnamefont {A.}~\bibnamefont {Moro}},\ }\bibfield
  {title} {\bibinfo {title} {On integrable conservation laws},\ }\href@noop {}
  {\bibfield  {journal} {\bibinfo  {journal} {Proc. R. Soc. A.}\ }\textbf
  {\bibinfo {volume} {471}} (\bibinfo {year} {2015}{\natexlab{b}})}\BibitemShut
  {NoStop}%
\bibitem [{\citenamefont {Erd{\"o}s}\ and\ \citenamefont
  {R\'enyi}(1959)}]{ErdosRenyi1959}%
  \BibitemOpen
  \bibfield  {author} {\bibinfo {author} {\bibfnamefont {P.}~\bibnamefont
  {Erd{\"o}s}}\ and\ \bibinfo {author} {\bibfnamefont {A.}~\bibnamefont
  {R\'enyi}},\ }\bibfield  {title} {\bibinfo {title} {On random graphs {I}},\
  }\href@noop {} {\bibfield  {journal} {\bibinfo  {journal} {Publicationes
  Mathematicae}\ }\textbf {\bibinfo {volume} {6}},\ \bibinfo {pages} {290}
  (\bibinfo {year} {1959})}\BibitemShut {NoStop}%
\bibitem [{\citenamefont {de~Bruno}(1857)}]{deBruno}%
  \BibitemOpen
  \bibfield  {author} {\bibinfo {author} {\bibfnamefont {F.}~\bibnamefont
  {de~Bruno}},\ }\bibfield  {title} {\bibinfo {title} {Note sur une nouvelle
  formule de calcul differentiel},\ }\href@noop {} {\bibfield  {journal}
  {\bibinfo  {journal} {Quartely J. of Pure and Appl. Math.}\ }\textbf
  {\bibinfo {volume} {1}},\ \bibinfo {pages} {359} (\bibinfo {year}
  {1857})}\BibitemShut {NoStop}%
\bibitem [{\citenamefont {Nittis}\ and\ \citenamefont
  {Moro}(2012)}]{DeNittisMoro2012}%
  \BibitemOpen
  \bibfield  {author} {\bibinfo {author} {\bibfnamefont {G.~D.}\ \bibnamefont
  {Nittis}}\ and\ \bibinfo {author} {\bibfnamefont {A.}~\bibnamefont {Moro}},\
  }\bibfield  {title} {\bibinfo {title} {Thermodynamic phase transitions and
  shock singularities},\ }\href@noop {} {\bibfield  {journal} {\bibinfo
  {journal} {Proc. R. Soc. A.}\ }\textbf {\bibinfo {volume} {468}},\ \bibinfo
  {pages} {701} (\bibinfo {year} {2012})}\BibitemShut {NoStop}%
\bibitem [{\citenamefont {Moro}(2014)}]{MoroAnnals2014}%
  \BibitemOpen
  \bibfield  {author} {\bibinfo {author} {\bibfnamefont {A.}~\bibnamefont
  {Moro}},\ }\bibfield  {title} {\bibinfo {title} {Shock dynamics of phase
  diagrams},\ }\href@noop {} {\bibfield  {journal} {\bibinfo  {journal} {Annals
  of Physics}\ }\textbf {\bibinfo {volume} {343}},\ \bibinfo {pages} {49}
  (\bibinfo {year} {2014})}\BibitemShut {NoStop}%
\bibitem [{\citenamefont {McKay}(2003)}]{McKay_book}%
  \BibitemOpen
  \bibfield  {author} {\bibinfo {author} {\bibfnamefont {D.~J.}\ \bibnamefont
  {McKay}},\ }\href@noop {} {\emph {\bibinfo {title} {{Information Theory,
  Inference, and Learning Algorithms}}}}\ (\bibinfo  {publisher} {{Cambridge
  University Press, New York}},\ \bibinfo {year} {2003})\BibitemShut {NoStop}%
\bibitem [{\citenamefont {Whitham}(1974)}]{Whitham_book}%
  \BibitemOpen
  \bibfield  {author} {\bibinfo {author} {\bibfnamefont {G.}~\bibnamefont
  {Whitham}},\ }\href@noop {} {\emph {\bibinfo {title} {{Linear and Nonlinear
  Waves}}}}\ (\bibinfo  {publisher} {{Wiley}},\ \bibinfo {year}
  {1974})\BibitemShut {NoStop}%
\bibitem [{\citenamefont {Whitney}(1955)}]{Whitney1955}%
  \BibitemOpen
  \bibfield  {author} {\bibinfo {author} {\bibfnamefont {H.}~\bibnamefont
  {Whitney}},\ }\bibfield  {title} {\bibinfo {title} {On singularities of
  mappings of euclidean spaces. i. mappings of the plane into the plane},\
  }\href@noop {} {\bibfield  {journal} {\bibinfo  {journal} {Annals of
  Mathematics}\ }\textbf {\bibinfo {volume} {62}},\ \bibinfo {pages} {374}
  (\bibinfo {year} {1955})}\BibitemShut {NoStop}%
\bibitem [{\citenamefont {Metropolis}\ \emph {et~al.}(1953)\citenamefont
  {Metropolis}, \citenamefont {Rosenbluth}, \citenamefont {Rosenbluth},
  \citenamefont {A.~Teller},\ and\ \citenamefont {Teller}}]{metropolis}%
  \BibitemOpen
  \bibfield  {author} {\bibinfo {author} {\bibfnamefont {N.}~\bibnamefont
  {Metropolis}}, \bibinfo {author} {\bibfnamefont {A.}~\bibnamefont
  {Rosenbluth}}, \bibinfo {author} {\bibfnamefont {M.}~\bibnamefont
  {Rosenbluth}}, \bibinfo {author} {\bibfnamefont {A.}~\bibnamefont
  {A.~Teller}},\ and\ \bibinfo {author} {\bibfnamefont {E.}~\bibnamefont
  {Teller}},\ }\bibfield  {title} {\bibinfo {title} {Equations of state
  calculations by fast computing machines},\ }\href@noop {} {\bibfield
  {journal} {\bibinfo  {journal} {J. Chem. Phys.}\ }\textbf {\bibinfo {volume}
  {21}},\ \bibinfo {pages} {1087} (\bibinfo {year} {1953})}\BibitemShut
  {NoStop}%
\bibitem [{\citenamefont {Hastings}(1970)}]{hastings}%
  \BibitemOpen
  \bibfield  {author} {\bibinfo {author} {\bibfnamefont {W.}~\bibnamefont
  {Hastings}},\ }\bibfield  {title} {\bibinfo {title} {Monte carlo sampling
  methods using markov chains and their applica-tion},\ }\href@noop {}
  {\bibfield  {journal} {\bibinfo  {journal} {Biometrika}\ }\textbf {\bibinfo
  {volume} {57}},\ \bibinfo {pages} {97} (\bibinfo {year} {1970})}\BibitemShut
  {NoStop}%
\bibitem [{\citenamefont {Brown}(1963)}]{Fuller1963}%
  \BibitemOpen
  \bibfield  {author} {\bibinfo {author} {\bibfnamefont {W.~F.}\ \bibnamefont
  {Brown}},\ }\bibfield  {title} {\bibinfo {title} {Thermal fluctuations of a
  single-domain particle},\ }\href@noop {} {\bibfield  {journal} {\bibinfo
  {journal} {Phys. Rev.}\ }\textbf {\bibinfo {volume} {130}},\ \bibinfo {pages}
  {1677} (\bibinfo {year} {1963})}\BibitemShut {NoStop}%
\bibitem [{\citenamefont {Park}\ and\ \citenamefont
  {Newman}(2005)}]{ParkNewman2005}%
  \BibitemOpen
  \bibfield  {author} {\bibinfo {author} {\bibfnamefont {J.}~\bibnamefont
  {Park}}\ and\ \bibinfo {author} {\bibfnamefont {M.~E.~J.}\ \bibnamefont
  {Newman}},\ }\bibfield  {title} {\bibinfo {title} {Solution for the
  properties of a clustered network},\ }\href@noop {} {\bibfield  {journal}
  {\bibinfo  {journal} {Phys. Rev. E}\ }\textbf {\bibinfo {volume} {70}},\
  \bibinfo {pages} {026136} (\bibinfo {year} {2005})}\BibitemShut {NoStop}%
\bibitem [{\citenamefont {Frank}\ and\ \citenamefont
  {Strauss}(1986)}]{FrankStrauss1986}%
  \BibitemOpen
  \bibfield  {author} {\bibinfo {author} {\bibfnamefont {O.}~\bibnamefont
  {Frank}}\ and\ \bibinfo {author} {\bibfnamefont {D.}~\bibnamefont
  {Strauss}},\ }\bibfield  {title} {\bibinfo {title} {Markov graphs},\
  }\href@noop {} {\bibfield  {journal} {\bibinfo  {journal} {Journal of the
  American Statistical Association}\ }\textbf {\bibinfo {volume} {81}},\
  \bibinfo {pages} {832} (\bibinfo {year} {1986})}\BibitemShut {NoStop}%
\bibitem [{\citenamefont {Annibale}\ and\ \citenamefont
  {Courtney}(2015)}]{AnnibaleCourtney2015}%
  \BibitemOpen
  \bibfield  {author} {\bibinfo {author} {\bibfnamefont {A.}~\bibnamefont
  {Annibale}}\ and\ \bibinfo {author} {\bibfnamefont {O.~T.}\ \bibnamefont
  {Courtney}},\ }\bibfield  {title} {\bibinfo {title} {The two-star model:
  exact solution in the sparse regime and condensation transition},\
  }\href@noop {} {\bibfield  {journal} {\bibinfo  {journal} {J. Phys. A: Math.
  Theor.}\ }\textbf {\bibinfo {volume} {48}},\ \bibinfo {pages} {365001}
  (\bibinfo {year} {2015})}\BibitemShut {NoStop}%
\bibitem [{\citenamefont {Agliari}\ \emph {et~al.}(2016)\citenamefont
  {Agliari}, \citenamefont {Barra}, \citenamefont {Schiavo},\ and\
  \citenamefont {Moro}}]{ABDM2016}%
  \BibitemOpen
  \bibfield  {author} {\bibinfo {author} {\bibfnamefont {E.}~\bibnamefont
  {Agliari}}, \bibinfo {author} {\bibfnamefont {A.}~\bibnamefont {Barra}},
  \bibinfo {author} {\bibfnamefont {L.~D.}\ \bibnamefont {Schiavo}},\ and\
  \bibinfo {author} {\bibfnamefont {A.}~\bibnamefont {Moro}},\ }\bibfield
  {title} {\bibinfo {title} {Complete integrability of information processing
  by biochemical reactions},\ }\href@noop {} {\bibfield  {journal} {\bibinfo
  {journal} {Sci. Rep.}\ }\textbf {\bibinfo {volume} {6}},\ \bibinfo {pages}
  {36314} (\bibinfo {year} {2016})}\BibitemShut {NoStop}%
\bibitem [{\citenamefont {Giglio}\ \emph {et~al.}(2018)\citenamefont {Giglio},
  \citenamefont {Matteis},\ and\ \citenamefont
  {Moro}}]{GiglioDeMatteisMoro2018}%
  \BibitemOpen
  \bibfield  {author} {\bibinfo {author} {\bibfnamefont {F.}~\bibnamefont
  {Giglio}}, \bibinfo {author} {\bibfnamefont {G.~D.}\ \bibnamefont
  {Matteis}},\ and\ \bibinfo {author} {\bibfnamefont {A.}~\bibnamefont
  {Moro}},\ }\bibfield  {title} {\bibinfo {title} {Exact equations of state for
  nematics},\ }\href@noop {} {\bibfield  {journal} {\bibinfo  {journal} {Annals
  of Physics}\ }\textbf {\bibinfo {volume} {396}},\ \bibinfo {pages} {386}
  (\bibinfo {year} {2018})}\BibitemShut {NoStop}%
\bibitem [{\citenamefont {Lorenzoni}\ and\ \citenamefont
  {Moro}(2018)}]{LorenzoniMoro2019}%
  \BibitemOpen
  \bibfield  {author} {\bibinfo {author} {\bibfnamefont {P.}~\bibnamefont
  {Lorenzoni}}\ and\ \bibinfo {author} {\bibfnamefont {A.}~\bibnamefont
  {Moro}},\ }\bibfield  {title} {\bibinfo {title} {Exact analysis of phase
  transitions in mean-field potts models},\ }\href@noop {} {\bibfield
  {journal} {\bibinfo  {journal} {Phys. Rev. E}\ }\textbf {\bibinfo {volume}
  {100}},\ \bibinfo {pages} {022103} (\bibinfo {year} {2018})}\BibitemShut
  {NoStop}%
\bibitem [{\citenamefont {Genovese}\ and\ \citenamefont
  {Barra}(2009)}]{GenoveseBarra2009}%
  \BibitemOpen
  \bibfield  {author} {\bibinfo {author} {\bibfnamefont {G.}~\bibnamefont
  {Genovese}}\ and\ \bibinfo {author} {\bibfnamefont {A.}~\bibnamefont
  {Barra}},\ }\bibfield  {title} {\bibinfo {title} {A mechanical approach to
  mean field spin models},\ }\href@noop {} {\bibfield  {journal} {\bibinfo
  {journal} {J. Math. Phys.}\ }\textbf {\bibinfo {volume} {50}},\ \bibinfo
  {pages} {053303} (\bibinfo {year} {2009})}\BibitemShut {NoStop}%
\bibitem [{\citenamefont {Benassi}\ and\ \citenamefont
  {Moro}(2020)}]{BenassiMoro2020}%
  \BibitemOpen
  \bibfield  {author} {\bibinfo {author} {\bibfnamefont {C.}~\bibnamefont
  {Benassi}}\ and\ \bibinfo {author} {\bibfnamefont {A.}~\bibnamefont {Moro}},\
  }\bibfield  {title} {\bibinfo {title} {Thermodynamic limit and dispersive
  regularization in matrix models},\ }\href@noop {} {\bibfield  {journal}
  {\bibinfo  {journal} {Phys. Rev. E}\ }\textbf {\bibinfo {volume} {101}}
  (\bibinfo {year} {2020})}\BibitemShut {NoStop}%
\bibitem [{\citenamefont {Benassi}\ \emph {et~al.}(2021)\citenamefont
  {Benassi}, \citenamefont {Dell'Atti},\ and\ \citenamefont
  {Moro}}]{BenassiDellAttiMoro2021}%
  \BibitemOpen
  \bibfield  {author} {\bibinfo {author} {\bibfnamefont {C.}~\bibnamefont
  {Benassi}}, \bibinfo {author} {\bibfnamefont {M.}~\bibnamefont {Dell'Atti}},\
  and\ \bibinfo {author} {\bibfnamefont {A.}~\bibnamefont {Moro}},\ }\bibfield
  {title} {\bibinfo {title} {Symmetric matrix ensemble and integrable
  hydrodynamic chains, preprint},\ }\href@noop {} {\bibfield  {journal}
  {\bibinfo  {journal} {arXiv::2101.10232}\ } (\bibinfo {year}
  {2021})}\BibitemShut {NoStop}%
\end{thebibliography}%

\end{document}